\documentclass[journal, 12pt, draftclsnofoot, onecolumn]{IEEEtran}

\IEEEoverridecommandlockouts

\usepackage{vwcol}
\usepackage{url}
\usepackage{graphicx}
\usepackage{psfrag}
\usepackage{amsmath}
\usepackage{amsthm}
\usepackage{amsfonts}
\usepackage{amssymb}
\usepackage{bm}
\usepackage{algorithm}
\usepackage{algorithmicx}

\usepackage{algpseudocode}
\usepackage{xcolor}
\usepackage{float}
\usepackage{bbm}
\usepackage{mathrsfs}
\usepackage{ulem}
\usepackage{bbold}
\usepackage[justification=centering]{caption}
\usepackage{subfigure}
\usepackage{cite}
\usepackage{hyperref} 
\usepackage{multicol}
\usepackage{tikz}
\usetikzlibrary{arrows.meta}
\usepackage{setspace}
\usepackage{color}
\usepackage[numbers,sort&compress]{natbib}
\usepackage{latexsym}
\usepackage{extarrows}
\usepackage{cancel}
\usepackage{lipsum}

\usepackage{pifont}
\usepackage{multirow}
\usepackage{indentfirst}
\usepackage{color}
\usepackage{ulem}

\usepackage{cuted}

\usepackage{tcolorbox}

\usepackage{makecell}

\usepackage{colortbl}
\usepackage{booktabs}
\usepackage{arydshln}

\makeatletter
\newenvironment{breakablealgorithm}
  {
   \begin{center}
     \refstepcounter{algorithm}
     \hrule height.8pt depth0pt \kern2pt
     \renewcommand{\caption}[2][\relax]{
       {\raggedright\textbf{\ALG@name~\thealgorithm} ##2\par}%
       \ifx\relax##1\relax 
         \addcontentsline{loa}{algorithm}{\protect\numberline{\thealgorithm}##2}%
       \else 
         \addcontentsline{loa}{algorithm}{\protect\numberline{\thealgorithm}##1}%
       \fi
       \kern2pt\hrule\kern2pt
     }
  }{
     \kern2pt\hrule\relax
   \end{center}
  }
\makeatother

\newcommand{\bs}{\boldsymbol}
\newcommand{\Sf}{\mathsf}
\newcommand{\TB}{\textbf}

\newcommand{\Eye}{\bs{I}}

\newcommand{\Trace}{\Sf{Tr}}
\newcommand{\Ts}{\Sf{T}}
\newcommand{\Cts}{\Sf{H}}

\newcommand{\Const}{\Sf{C}}

\newcommand{\Bs}{\backslash}

\newcommand{\Normal}{\Sf{N}}
\newcommand{\NormalC}{\Sf{N}_{\Complex}}

\newcommand{\dd}{\Sf{d}}

\newcommand{\Proj}{\Sf{Proj}}
\newcommand{\Diag}{\Sf{D}}
\newcommand{\diag}{\Sf{d}}

\newcommand{\Mean}{\mathbb{E}}
\newcommand{\Var}{\mathbb{V}}

\newcommand{\Ext}{\Sf{Ext}}

\newcommand{\Complex}{\mathbb{C}}

\newcommand{\Lex}{\Sf{Lex}}

\newcommand{\Pex}{\Sf{Pex}}
\newcommand{\Ez}{\Sf{Ez}}

\newcommand{\Eig}{\Sf{eig}}

\newcommand{\EXP}{\Sf{e}}

\newcommand{\Mbar}{\underline{m}}

\newcommand{\ArgMin}{\mathop{\operatorname{argmin}}}
\newcommand{\ArgMax}{\mathop{\operatorname{argmax}}}

\newcommand{\KL}{\Sf{KL}}

\newcounter{magicrownumbers}

\allowdisplaybreaks[4]

\ifCLASSINFOpdf

\else

\fi

\begin{document}

\title{
Quadratic Message Passing for Generalized Quadratic Equations Model
}

\author{
Huimin Zhu*
\thanks{
H. Zhu is with Guangzhou University of Chinese Medicine, Guangzhou 510006, China e-mail: (hm\_zhu@gzucm.edu.cn).
}
}

\maketitle

\begin{abstract}

For approximate inference in the generalized quadratic equations model,
many state-of-the-art algorithms lack any prior knowledge of the target signal structure, exhibits slow convergence, and can not handle any analytic prior knowledge of the target signal structure.
So, this paper proposes a new algorithm, Quadratic Message passing (QMP).
QMP has a complexity as low as $O(N^{3})$.
The SE derived for QMP can capture precisely the per-iteration behavior of the simulated algorithm. 
Simulation results confirm QMP outperforms many state-of-the-art algorithms.

\end{abstract}


\IEEEpeerreviewmaketitle

\section{Introduction}

We are interested in recovering the target signal 
$\tilde{\bs{x}} \in \Complex^{N \times 1}$
from the quadratic measurements
$\{y_{i}\}_{i = 1, \cdots, M}$
given as
\begin{align}
y_{i}
= &
\Sf{Q}(z_{i})
, \qquad
z_{i}
\triangleq
\tilde{\bs{x}}^{\Cts} \bs{A}_{i} \tilde{\bs{x}}
,\label{Eq:System_Model}
\end{align}
where
$\{\bs{A}_{i}\}_{i = 1, \cdots, M} \in \Complex^{N \times N}$
are known measurement matrices and
$\Sf{Q}(\cdot)$
is the deterministic and separable likelihood probability function.

The generalized quadratic equations model is applicable to various scenarios:
\begin{itemize}

\item

\TB{Sparsity based sub-wavelength imaging:}
Sub-wavelength optical images captured under partially spatially incoherent light can be reconstructed from either their far-field or blurred images, given prior knowledge that the image is sparse.
The reconstruction method relies on sparsity-based sub-wavelength imaging, recently demonstrated.
This new algorithmic methodology, known as quadratic compressed sensing, can be extended to various other problems involving information recovery from partially correlated measurements, even in cases where the correlation function exhibits local dependencies \cite{shechtman2011sparsity};

\item

\TB{Quadratic (0, 1)-problem}:
Optimizing a quadratic function over a hypercube is a fundamental problem in discrete optimization.
However, this problem is NP-hard, and solving general moderate-sized instances to optimality remains a significant computational challenge.
Quadratic programming over hypercube vertices is found in various equivalent formulations in the literature \cite{helmberg1998solving}.

\end{itemize}

Various algorithms have been proposed to tackle this problem.
Error-reduction methods \cite{gerchberg1994practical, fienup1982phase} have shown promising empirical performance but lack strong theoretical guarantees.
Convex relaxation techniques \cite{candes2013phaselift, candes2014solving} have been developed.
These methods offer performance guarantees, and corresponding algorithms have been devised accordingly.

The authors proposed to achieve a globally optimal solution through a standard framework involving spectral initialization and iterative Wirtinger flow (WF) updates.
And they demonstrated that when the number of the quadratic measurements exceeds the target signal length by some sufficiently large constant, the target signal can be recovered with high probability, albeit subject to a global phase shift.
However, the WF algorithm only operates in the high-dimensional case ($M > N$), lacking any prior knowledge of the target signal structure, and exhibits slow convergence.

Furthermore, the authors introduced a non-convex algorithm called Thresholded Wirtinger Flow (TWF) to address the low-dimensional scenario ($M < N$), leveraging available prior knowledge or structural assumptions about the target signal.
The TWF algorithm is capable of provably reconstructing the target signal, a feat that many existing non-convex methods struggle to achieve in terms of (global) convergence guarantees of this nature.
But, the TWF algorithm can not handle any analytic prior knowledge of the target signal structure.

For addressing the generalized quadratic equations model (\ref{Eq:System_Model}), we apply the convenient framework of Bayesian inference, in which we are especially interested in the minimum mean square error (MMSE) estimator whose output is denoted as
\begin{align}
\tilde{\bs{x}}^{\text{MMSE}}
= & 
\int{\dd \tilde{\bs{x}}}\,
\tilde{\bs{x}} p(\tilde{\bs{x}} | \bs{y}; 1)
,\label{Eq:MMSE_estimator}
\end{align}
and maximum a posteriori (MAP) estimator whose output is referred as
\begin{align}
\tilde{\bs{x}}^{\text{MAP}}
= &
\ArgMax_{\tilde{\bs{x}}} p(\tilde{\bs{x}} | \bs{y}; +\infty)
,\label{Eq:MAP_estimator}
\end{align}
where both estimators require the marginal distribution
$p(\tilde{\bs{x}} | \bs{y}; \beta)$
, which is the marginalization of the posterior distribution
$p(\tilde{\bs{x}} | \bs{y}; \beta)$
denoted as
\begin{align}
p(\tilde{\bs{x}} | \bs{y}; \beta)
\propto &
\int{\dd \bs{z}}\,
\prod_{i = 1}^{M}{}[
	p^{\beta}(y_{i} | z_{i})
	\delta(
		z_{i}
		-
		\tilde{\bs{x}}^{\Cts} \bs{A}_{i} \tilde{\bs{x}}
	)
]
p^{\beta}(\tilde{\bs{x}})
\label{Eq:Joint_Probability}\\
\propto &
\int{
	\dd \bs{z}
	\dd \check{\bs{x}}_{1}
	\cdots
	\dd \check{\bs{x}}_{M}
	\dd \bs{x}
}\,
[
	\prod_{i = 1}^{M}{}
	p^{\beta}(y_{i} | z_{i})
	\delta(
		z_{i}
		-
		\check{\bs{x}}_{i}^{\Cts} \bs{A}_{i} \bs{x}
	)
	\delta(\check{\bs{x}}_{i} - \bs{x})
]
\delta(\bs{x} - \tilde{\bs{x}})
p^{\beta}(\tilde{\bs{x}})
.\label{Eq:Joint_Probability_Reformula}
\end{align}
where
$
p(\tilde{\bs{x}})
\triangleq
\prod_{i = 1}^{N} p(\tilde{x}_{i})
$
and $\beta$ is introduced to unify the MMSE ($\beta = 1$) and MAP ($
\beta = + \infty
$) estimators. 

\TB{Organization:}
The remainder is organized as follows.
In section II, we propose a algorithm based on hybrid message passing for the generalized quadratic equations model, which is called as \textit{Quadratic Message Passing} (QMP).
In section III, we present the SE analysis of QMP. 
In section IV, we detail the applications of QMP, and present the simulated results to verify the performance of QMP.

In summary, this paper contributes in two aspects:
(i) We propose QMP, a message passing algorithm, for approximate inference under the generalized quadratic equations model.
The computational complexity of QMP is on the order of $O(N^{3})$, with $N$ being the problem size. 
We empirically show QMP significantly outperforms the state-of-the-art algorithms of WF and TWF.
(ii) We present an SE for QMP, which is shown to capture precisely the dynamical behaviors of the algorithm.

\TB{Notations:}
Throughout this paper, we adopt the following notation conventions:
non-bold lowercase letters (e.g., $m$ and $v$) represent scalars, bold lowercase letters (e.g., $\bs{m}$ and $\bs{v}$) denote column vectors, bold lowercase letters with a vector arrow (e.g., $\vec{\bs{m}}$ and $\vec{\bs{v}}$) indicate row vectors, and capital letters (e.g., $\bs{V}$ and $\bs{C}$) signify matrices.
$\EXP(\cdot)$ represents an exponential operator.
$\delta(\cdot)$ denotes a Dirac delta function.
$\Normal[\bs{m}, \bs{C}]$ denotes a Gaussian distribution with mean $\bs{m}$ and covariance $\bs{C}$, defined as
$
\Normal[\bs{x} | \bs{m}, \bs{C}]
\triangleq
|2 \pi \bs{C}|^{- \frac{1}{2}}
\EXP[
	- \frac{1}{2}
	(\bs{x} - \bs{m})^{\Ts} \bs{C}^{-1} (\bs{x}-\bs{m})
]
$.
For any matrix $\bs{A}$, $A^{(i, j)}$ represents the element at the $i$-th row and $j$-th column of $\bs{A}$.
$\bs{A}^\Ts$ denotes the transpose of matrix $\bs{A}$.
$\bs{A}^{\Cts}$ represents the conjugate transpose of $\bs{A}$.
$\Trace(\bs{A})$ stands for the trace of matrix $\bs{A}$.
$\|\bs{A}\|_{\mathrm{F}}$ denotes the Frobenius norm.
$\bs{1}_{N}$ is a column vector of size $N$ consisting of all ones.
$\Diag(\bs{v})$ is a diagonal matrix with diagonal elements equal to the elements of vector $\bs{v}$.
$\diag(\bs{C})$ is a diagonal operator, returning a $N$-dimensional column vector containing the diagonal elements of matrix $\bs{C}$.
$\odot$ and $\oslash$ denote element-wise vector multiplication and division, respectively.
$\Const$ denotes a constant.

\section{The QMP Algorithm}

To approximate the marginal of the high-dimensional posterior joint
$p(\tilde{\bs{x}} | \bs{y}; \beta)$
, we resort to message passing (a.k.a. belief propagation) a deterministic method that performs on the  factor graph.
A factor graph is a bipartite probabilistic graph model that represents explicitly the algebraic structure of the joint density function.
In the factor graph, circles are variable nodes, representing the hidden/latent random variables, while squares are factor nodes, representing the transitional/conditional probability density that relates the random variable(s). 
An edge, in either direction, connecting a variable node with a factor node is associated with a  message.
The messages traverse through the graph following some rules appropriately designed, and are  able to offer/approximate the desired marginal result.

Two ingredients are key to the development of a message passing algorithm: 
1) a set of message updating rules that balance the approximation precision and the computational complexity; 
2) a suitable factor graph that facilitates the message updating. 
These two ingredients complement each other and are an intrinsic whole. 

\begin{figure}[!t]
\centering
\includegraphics[width=0.7\textwidth]{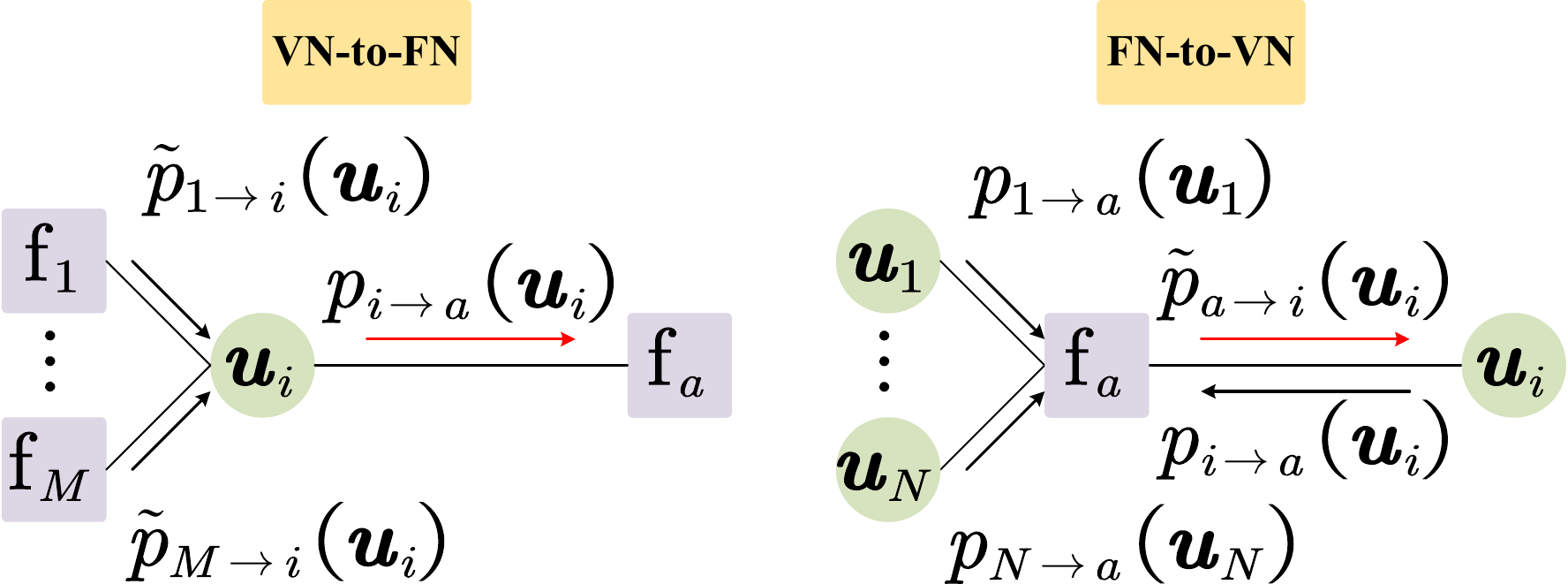}
\caption{A factor graph for explaining message passing rules}
\label{Fig:FG_GaBP_EP}
\end{figure}

\subsection{Hybrid Message Updating and Vector-Form Factor Graph}
Prevailing in the literature are three types of message passing/updating:
i) standard belief propagation (BP),
ii) Gaussian belief propagation (GaBP, also known as the relaxed BP),
iii) expectation propagation (EP, also known as single-loop expectation consistent) \cite{bishop2006pattern}.
Below we review the three briefly by using Fig. \ref{Fig:FG_GaBP_EP}, and will elaborate on the discussion of their pros and cons.

i) \textbf{Standard BP} updates the messages according to the following rules:
\begin{align}
p_{i \rightarrow a}(\bs{u}_{i})
= &
\prod_{b \neq a}^{M}{}
\tilde{p}_{b \rightarrow i}(\bs{u}_{i})
,\label{eq:BP1}\\
\tilde{p}_{a \rightarrow i}(\bs{u}_{i})
= &
\tilde{\Sf{f}}_{a \rightarrow i}(\bs{u}_{i})
,\label{eq:BP2}
\end{align}
where
$
\tilde{\Sf{f}}_{a \rightarrow i}(\bs{u}_{i})
\triangleq 
\int{}\, \Sf{f}_{a}(\bs{u}_{1}, \cdots, \bs{u}_{N})
\prod_{j \neq i}^{N}{}
[
	p_{j \rightarrow a}(\bs{u}_{j}) \dd \bs{u}_{j}
]
$.
Standard BP, while performing on a tree factor graph, is provably exact for the computation of marginals.
However, on graphs that contain loops, it has no guarantee of exactness and convergence in general. Fortunately, for graphs locally tree like, BP can be proved to be effective \cite{mezard2009information}.
The intuition here is: being a local algorithm, BP should be successful whenever the underlying graph is locally a tree.
Such factor graphs appear frequently, for instance, in error-correcting codes, where BP was long known to be powerful.
The biggest problem with BP here is the exact evaluation of a high-dimensional integral
$\tilde{\Sf{f}}_{a \rightarrow i}(\bs{u}_{i})$. 
In total, there are $M \times N$ such integrals, with $M$ and $N$ both being sufficiently large.

ii) \textbf{EP} uniformly approximates all messages in both directions as Gaussian densities, i.e., 
\begin{align}
p_{i \rightarrow a}(\bs{u}_{i})
\propto &
\prod_{b \neq a}^{M}{}
\tilde{p}_{b \rightarrow i}(\bs{u}_{i})
\approx
\Normal(
	\bs{u}_{i} |
	\bs{m}_{i \rightarrow a},
	\bs{v}_{i \rightarrow a}
)
,\label{eq:EP1}\\
\tilde{p}_{a \rightarrow i}(\bs{u}_{i})
\propto &
\frac{
	\Proj[
		\tilde{\Sf{f}}_{a \rightarrow i}(\bs{u}_{i})
		p_{i \rightarrow a}(\bs{u}_{i})
	]
}{
	p_{i \rightarrow a}(\bs{u}_{i})
}
\approx
\Normal(
	\bs{u}_{i} |
	\tilde{\bs{m}}_{a \rightarrow i},
	\tilde{\bs{v}}_{a \rightarrow i}
)
,\label{eq:EP2}
\end{align}
where $\bs{m}$ and $\tilde{\bs{m}}$ are the means, and $\bs{v}$ and $\tilde{\bs{v}}$ are the diagonal elements of the (diagonal) covariance matrices.
The projection operator $\Proj[g(\bs{x})]$  approximates a generic function $g(\bs{x})$ by a Gaussian probability density function (PDF) that minimizes the Kullback-Leibler divergence (KLD) over all possible normal distributions $\Normal$:
\begin{align}
\Proj[g(\bs{x})] 
\triangleq &
\ArgMin_{\tilde{g} \in \Normal}
\KL[g \| \tilde{g}]
= 
\Normal(\bs{x} | \bs{m}_{g}, \bs{v}_{g}).
\end{align}

The KLD, defined as 
$
\KL[g \| \tilde{g}]
\triangleq 
\int{\dd \bs{x}}\,
g(\bs{x}) \log[
	\frac{
		g(\bs{x})
	}{
		\tilde{g}(\bs{x})
	}
]
$,
is in general difficult to optimize; however, in case $\tilde{g}$ is in the exponential family, the minimization of $\KL[g \| \tilde{g}]$ becomes precisely solvable, and the optimal condition is 
$
\Mean_{\bs{x} \sim g}[\phi(\bs{x})]
=
\Mean_{\bs{x} \sim \tilde{g}} [\phi(\bs{x})]$, where $\phi(X)$ is the so-called \emph{sufficient statistics} of the very exponential family $\tilde{g}$ belongs to \cite{bishop2006pattern}.
For EP, the above optimal condition is identical to the matching of moments (indeed, cumulants) between the true density $g$ and its Gaussian approximate $\tilde{g}$, i.e., 
\begin{align}
\bs{m}_{g}
= &
\frac{
	\int{\dd \bs{x}}\, \bs{x} g(\bs{x})
}{
	\int{\dd \bs{x}}\, g(\bs{x})
}
,\label{eq:GaBP_1}\\
\bs{v}_{g}
= &
\frac{
	\diag[
		\int{\dd \bs{x}}\,
		(\bs{x} - \bs{m}_{g})
		(\bs{x} - \bs{m}_{g})^{\Ts} g(\bs{x})
	]
}{
	\int{\dd \bs{x}}\, g(\bs{x})
}
.\label{eq:GaBP_2}
\end{align}
The projection in EP is therefore called \emph{moment matching}.
It is also worthy of noting, in (\ref{eq:EP2}), EP multiplies $\tilde{\Sf{f}}_{a \to i}$ with $p_{i \to a}$ before the projection but removes it immediately after. 
This is to force the projected term to be exponential-family so that its evaluation of mean and variance may be sped up.
In this way, EP can avoid the intractability problem faced by BP, since all messages are uniformly Gaussian:
Eq. (\ref{eq:EP2}) is Gaussian, as long as $p_{i \to a}$ is initialized as Gaussian; 
Eq. (\ref{eq:EP1}) is also Gaussian, as a result of the Gaussian reproducing property \cite{rasmussen2003gaussian}, which says the product of two Gaussian PDFs is still (proportional to)  Gaussian, i.e.,
$
\Normal(x | a, A) \Normal(x|b, B)
=
\Normal(0 | a - b, A + B)
\Normal[
	x |
	(\frac{1}{A} + \frac{1}{B})^{- 1}
	(\frac{a}{A} + \frac{b}{B}),
	(\frac{1}{A} + \frac{1}{B})^{- 1}
]
$.

iii) \textbf{GaBP}, unlike EP, which requires all messages be formally Gaussian, allows its messages to take a relaxed Gaussian-like form.
To be specific, GaBP has its messages updated as follows, where the two matrices,
$\bs{\Lambda}_{i \to a}$
and
$\tilde{\bs{\Lambda}}_{a \to i}$,
can be rank-deficient
\begin{align}
p_{i \rightarrow a}(\bs{u}_{i})
= &
\prod_{b \neq a}^{M}{}
\tilde{p}_{b \rightarrow i}(\bs{u}_{i})
\approx
\EXP(
	- \frac{1}{2}
	\bs{u}^{\Ts}_{i} \bs{\Lambda}_{i \rightarrow a}
	\bs{u}_{i}
	+ \bs{u}^{\Ts}_{i} \bs{b}_{i\rightarrow a}
)
,\label{eq:GaBP1}\\
\tilde{p}_{a \rightarrow i}(\bs{u}_{i})
= &
\Proj[
	\tilde{\Sf{f}}_{a \rightarrow i}(\bs{u}_{i})
]
\approx 
\EXP(
	- \frac{1}{2}
	\bs{u}^{\Ts}_{i}
	\tilde{\bs{\Lambda}}_{a \rightarrow i}
	\bs{u}_{i}
	+ \bs{u}^{\Ts}_{i} \tilde{\bs{b}}_{a \rightarrow i}
)
,\label{eq:GaBP2}
\end{align}
where
\begin{align*}
\Proj[g(\bs{x})] 
\triangleq &
\ArgMin_{\tilde{g} \in \mathcal{G}} \KL[ g \| \tilde{g}]
=
\EXP(
	- \frac{1}{2}
	\bs{x}^{\Ts} \bs{\Lambda}_{g}
	\bs{x}
	+ \bs{x}^{\Ts} \bs{b}_{g}
)
,
\end{align*}
$
\mathcal{G}
\triangleq
\{  
	\EXP(
		- \frac{1}{2}
		\bs{x}^{\Ts} \bs{\Lambda} \bs{x}
		+ \bs{x}^{\Ts} \bs{b}
	), 
	\forall \bs{\Lambda}, \forall \bs{b}
\}
$,
and the optimal condition for the KLD minimization remains unchanged, which is
$
\Mean_{\bs{x} \sim g}[\phi(\bs{x})]
=
\Mean_{\bs{x} \sim \tilde{g}}[\phi(\bs{x})]
$,
while the sufficient statistic is still
$
\phi(\bs{x})
=
[
	\bs{x}^{\Ts},
	\mathrm{vec}^{\Ts}(\bs{x} \bs{x}^{\Ts})
]^{\Ts}
$.
However, solving the optimal condition for the desired parameters $\bs{\Lambda}_{g}$ and $\bs{b}_{g}$ is not as easy.

In summary, we present the posterior distribution in (\ref{Eq:Joint_Probability_Reformula}) as a factor graph with vector-valued nodes, described as in Fig. \ref{Fig:QMP}.
In own factor graph, we can consider the blue block as a "false" factor node
$
\delta(
	z_{i}
	-
	\tilde{\bs{x}}^{\Cts} \bs{A}_{i} \tilde{\bs{x}}
)
$
, and apply the EP rules for all messages outside the "false" factor node. For the messages inside the "false" factor node, we can find that the EP approximation of messages from
$
\delta(
	z_{i}
	-
	\check{\bs{x}}_{i}^{\Cts} \bs{A}_{i} \bs{x}
)
$
to
$\check{\bs{x}}_{i}^{\Cts}$
and
$\bs{x}$
is so aggressive, because it simply approximates a Gaussian distribution with rank deficient covariance matrix to one with full rank covariance matrix. So we correspondingly apply the GaBP rules inside the "false" factor node to more accurately approximate as the general Gaussian form for remaining the rank deficient convariance matrix structure. 

\subsection{Scheduling of the Iterative Message Passing}

In this part, we propose the quadratic message passing algorithm (QMP), whose basic building blocks are GaBP and EP.

\TB{Schedule of QMP}:
Following the above operations, we can obtain all messages of the QMP in Table \ref{Tab:QMP}. 
Now we start to analyze the iterative steps of QMP as: 

\begin{figure}
\centering
\includegraphics[width=0.7\textwidth]{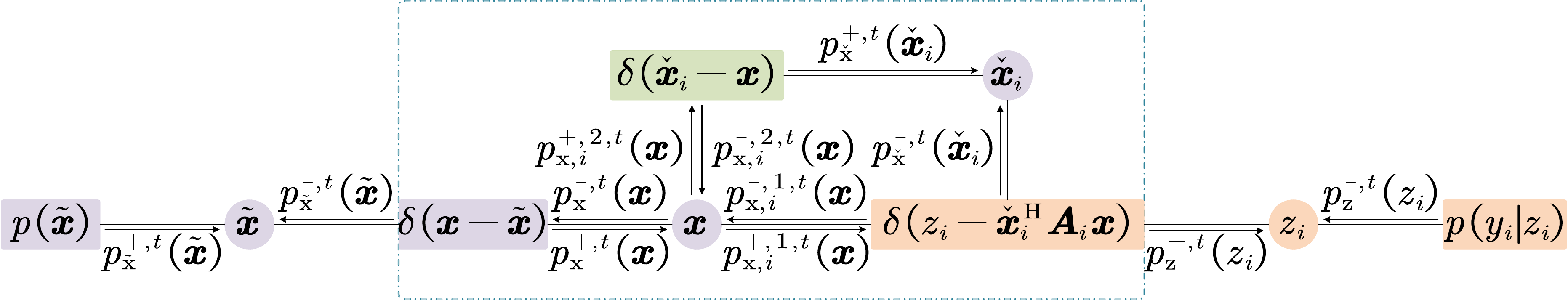}
\caption{The factor graph of QMP}
\label{Fig:QMP}
\end{figure}

\begin{table}[!t]
\centering
\caption{Passing messages in the factor graph}
\label{Tab:QMP}
\begin{tabular}{| c | c |}\hline
$
p_{\tilde{\Sf{x}}}^{+, t}(\tilde{\bs{x}})
$ & EP message from $
p(\tilde{\bs{x}})
$ to $
\tilde{\bs{x}}
$ \\
$
p_{\tilde{\Sf{x}}}^{-, t}(\tilde{\bs{x}})
$ & EP message from $
\delta(\bs{x} - \tilde{\bs{x}})
$ to $
\tilde{\bs{x}}
$ \\
$
p_{\Sf{x}}^{+, t}(\bs{x})
$ & GaBP message from $
\delta(\bs{x} - \tilde{\bs{x}})
$ to $
\bs{x}
$ \\
$
p_{\Sf{x}}^{-, t}(\bs{x})
$ & GaBP message from $
\bs{x}
$ to $
\delta(\bs{x} - \tilde{\bs{x}})
$ \\
$
p_{\Sf{x}, i}^{+, 1, t}(\bs{x})
$ & GaBP message from $
\bs{x}
$ to $
\delta(
	z_{i}
	-
	\check{\bs{x}}_{i}^{\Cts}
	\bs{A}_{i}
	\bs{x}
)
$ \\
$
p_{\Sf{x}, i}^{-, 1, t}(\bs{x})
$ & GaBP message from $
\delta(
	z_{i}
	-
	\check{\bs{x}}_{i}^{\Cts}
	\bs{A}_{i}
	\bs{x}
)
$ to $
\bs{x}
$ \\
$
p_{\Sf{x},i}^{+, 2, t}(\bs{x})
$ & GaBP message from $
\bs{x}
$ to $
\delta(\check{\bs{x}}_{i} - \bs{x})
$ \\
$
p_{\Sf{x}, i}^{-, 2, t}(\bs{x})
$ & GaBP message from $
\delta(\check{\bs{x}}_{i} - \bs{x})
$ to $
\bs{x}
$ \\
$
p_{\check{\Sf{x}}}^{+, t}(\check{\bs{x}}_{i})
$ & GaBP message from $
\delta(\check{\bs{x}}_{i} - \bs{x})
$ to $
\check{\bs{x}}_{i}
$ \\
$
p_{\check{\Sf{x}}}^{-, t}(\check{\bs{x}}_{i})
$ & GaBP message from $
\delta(
	z_{i}
	-
	\check{\bs{x}}_{i}^{\Cts}
	\bs{A}_{i}
	\bs{x}
)
$ to $
\check{\bs{x}}_{i}
$ \\
$
p_{\Sf{z}}^{+, t}(z_{i})
$ & EP message from $
\delta(
	z_{i}
	-
	\check{\bs{x}}_{i}^{\Cts}
	\bs{A}_{i}
	\bs{x}
)
$ to $
z_{i}
$ \\
$
p_{\Sf{z}}^{-, t}(z_{i})
$ & EP message from $
p(y_{i} | z_{i})
$ to $
z_{i}
$\\\hline
\end{tabular}
\end{table}

S1): Initialize all forward messages as Gaussian form densities;

S2): Backward passing, the messages delivered from right to left denote as
\begin{align}
p_{\Sf{z}}^{-, t}(z_{i})
= &
\frac{
	\Proj[p_{\Sf{z}}^{+, t}(z_{i}) p(y_{i} | z_{i})]
}{
	p_{\Sf{z}}^{+, t}(z_{i})
}
,\label{Eq:P_s_z}\\
p_{\Sf{x},i}^{-, 1, t}(\bs{x})
= &
\int{\dd z_{i} \dd \check{\bs{x}}_{i}}\,
\delta(z_{i} - \check{\bs{x}}_{i}^{\Cts} \bs{A}_{i} \bs{x})
p_{\Sf{z}}^{-, t}(z_{i})
p^{+, t}_{\check{\Sf{x}}}(\check{\bs{x}}_{i})
,\label{Eq:P_s1_xi}\\
p^{-, t}_{\check{\Sf{x}}}(\check{\bs{x}}_{i})
= &
\int{\dd z_{i} \dd \bs{x}}\,
\delta(z_{i} - \check{\bs{x}}_{i}^{\Cts} \bs{A}_{i} \bs{x})
p_{\Sf{z}}^{-, t}(z_{i})
p_{\Sf{x}, i}^{+, 1, t}(\bs{x})
,\label{Eq:P_s_cx}\\
p_{\Sf{x},i}^{-, 2, t}(\bs{x})
= &
\int{\dd \check{\bs{x}}_{i}}\,
\delta(\check{\bs{x}}_{i} - \bs{x})
p_{\check{\Sf{x}}}^{-, t}(\check{\bs{x}}_{i})
,\label{Eq:P_s2_xi}\\
p_{\Sf{x}}^{-, t}(\bs{x})
= &
\prod_{i = 1}^{M}{}[
	p_{\Sf{x}, i}^{-, 1, t}(\bs{x})
	p_{\Sf{x}, i}^{-, 2, t}(\bs{x})
]
,\label{Eq:P_s_x}\\
p_{\tilde{\Sf{x}}}^{-, t}(\tilde{\bs{x}})
= &
\frac{
	\Proj[
		p_{\tilde{\Sf{x}}}^{+, t}(\tilde{\bs{x}})
		f_{\tilde{\Sf{x}}}^{-, t}(\tilde{\bs{x}})
	]
}{
	p_{\tilde{\Sf{x}}}^{+, t}(\tilde{\bs{x}})
}
,\label{Eq:P_s_tx}
\end{align}
where
\begin{align}
f_{\tilde{\Sf{x}}}^{-, t}(\tilde{\bs{x}})
= &
\int{\dd \bs{x}}\,
\delta(\bs{x} - \tilde{\bs{x}})
p_{\Sf{x}}^{-, t}(\bs{x})
;\label{Eq:f_s_tx}
\end{align}

S3): Forward passing, the messages delivered from left to right represent as
\begin{align}
p_{\tilde{\Sf{x}}}^{+, t + 1}(\tilde{\bs{x}})
= &
\frac{
	\Proj[
		p_{\tilde{\Sf{x}}}^{-, t}(\tilde{\bs{x}})
		p(\tilde{\bs{x}})
	]
}{
	p_{\tilde{\Sf{x}}}^{-, t}(\tilde{\bs{x}})
}
,\label{Eq:P_p_tx}\\
p_{\Sf{x}}^{+, t + 1}(\bs{x})
= &
\int{\dd \tilde{\bs{x}}}\,
\delta(\bs{x} - \tilde{\bs{x}})
p_{\tilde{\Sf{x}}}^{+, t + 1}(\tilde{\bs{x}})
,\label{Eq:P_p_x}\\
p_{\Sf{x}, i}^{+, 1, t + 1}(\bs{x})
= &
p_{\Sf{x}}^{+, t + 1}(\bs{x})
[
	\prod_{k \neq i}^{M}{}
	p_{\Sf{x}, k}^{-, 1, t}(\bs{x})
]
[
	\prod_{k = 1}^{M}{}
	p_{\Sf{x}, k}^{-, 2, t}(\bs{x})
]
,\label{Eq:P_p1_xi}\\
p_{\Sf{x}, i}^{+, 2, t + 1}(\bs{x})
= &
p_{\Sf{x}}^{+, t + 1}(\bs{x})
[
	\prod_{k = 1}^{M}{}
	p_{\Sf{x}, k}^{-, 1, t}(\bs{x})
]
[
	\prod_{k \neq i}^{M}{}
	p_{\Sf{x}, k}^{-, 2, t}(\bs{x})
]
,\label{Eq:P_p2_xi}\\
p_{\check{\Sf{x}}}^{+, t + 1}(\check{\bs{x}}_{i})
= &
\int{\dd \bs{x}}\,
\delta(\check{\bs{x}}_{i} - \bs{x})
p_{\Sf{x}, i}^{+, 2, t + 1}(\bs{x})
,\label{Eq:P_p_cx}\\
p_{\Sf{z}}^{+, t + 1}(z_{i})
= &
\int{\dd \check{\bs{x}}_{i} \dd \bs{x}}\,
\delta(z_{i} - \check{\bs{x}}_{i}^{\Cts} \bs{A}_{i} \bs{x})
p_{\check{\Sf{x}}}^{+, t + 1}(\check{\bs{x}}_{i})
p_{\Sf{x}, i}^{+, 1, t + 1}(\bs{x})
.\label{Eq:P_p_z}
\end{align}

For \TB{S1}, we initialize $
p_{\check{\Sf{x}}}^{+, t}(\check{\bs{x}}_{i})
$ and $
p_{\Sf{x}, i}^{+, 1, t}(\bs{x})
$ by the general $N$-dimensional Gaussian forms as
\begin{align}
p_{\check{\Sf{x}}}^{+, t}(\check{\bs{x}}_{i})
\propto &
\EXP(
	- \frac{1}{2}
	\check{\bs{x}}_{i}^{\Cts}
	\bs{\Lambda}_{\check{\Sf{x}}, i}^{+, t}
	\check{\bs{x}}_{i}
	+ \check{\bs{x}}_{i}^{\Cts}
	\bs{b}_{\check{\Sf{x}}, i}^{+, t}
)
,\label{Eq:Init_P_p_cx}\\
p_{\Sf{x}, i}^{+, 1, t}(\bs{x})
\propto &
\EXP(
	- \frac{1}{2}
	\bs{x}^{\Cts}
	\bs{\Lambda}_{\Sf{x}, i}^{+, 1, t}
	\bs{x}
	+ \bs{x}^{\Cts}
	\bs{b}_{\Sf{x},i}^{+,1,t}
)
,\label{Eq:Init_P_p1_xi}
\end{align}
and parameterize the remaining forward passing messages as
\begin{align}
p_{\Sf{z}}^{+, t}(\bs{z})
= &
\NormalC[
	\bs{z} |
	\bs{m}_{\Sf{z}}^{+, t}, \bs{v}_{\Sf{z}}^{+, t}
]
,\label{Eq:Init_P_p_z}\\
p_{\tilde{\Sf{x}}}^{+, t}(\tilde{\bs{x}})
= &
\NormalC[
	\tilde{\bs{x}} |
	\bs{m}_{\tilde{\Sf{x}}}^{+, t},
	\bs{v}_{\tilde{\Sf{x}}}^{+, t}
].
\label{Eq:Init_P_p_tx}
\end{align}

For \TB{S2}, the procedure of deriving all backward passing messages is shown as below.

\TB{Step 2.1}: We denote the projection function in the numerator of \eqref{Eq:P_s_z} as $
\NormalC[
	\bs{z} |
	\hat{\bs{m}}_{\Sf{z}}^{-, t},
	\hat{\bs{v}}_{\Sf{z}}^{-, t}
]
$, where $
\hat{\bs{m}}_{\Sf{z}}^{-, t}
=
\Mean[
	\bs{z} |
	\bs{m}_{\Sf{z}}^{+, t},
	\bs{v}_{\Sf{z}}^{+, t}
]
$, $
\hat{\bs{v}}_{\Sf{z}}^{-, t}
=
\Var[
	\bs{z} |
	\bs{m}_{\Sf{z}}^{+, t}, \bs{v}_{\Sf{z}}^{+, t}
]
$ are the mean and variance of the approximate posterior distribution $
p(\bs{z} | \bs{y})
=
\frac{
	p(\bs{y} | \bs{z})
	\NormalC[
		\bs{z} |
		\bs{m}_{\Sf{z}}^{+, t},
		\bs{v}_{\Sf{z}}^{+, t}
	]
}{
	\int{\dd \bs{z}}\,
	p(\bs{y} | \bs{z})
	\NormalC[
		\bs{z} |
		\bs{m}_{\Sf{z}}^{+, t},	
		\bs{v}_{\Sf{z}}^{+, t}
	]
}
$, respectively. Then we compute the entirety of \eqref{Eq:P_s_z} via the Gaussian reproduction property as
\begin{align}
p_{\Sf{z}}^{-, t}(\bs{z})
= &
\NormalC[
	\bs{z} |
	\bs{m}_{\Sf{z}}^{-, t},
	\bs{v}_{\Sf{z}}^{-, t}
]
,\label{Eq:Update_P_s_z}
\end{align}
where 
\begin{align}
\bs{v}_{\Sf{z}}^{-, t}
= &
\bs{1} \oslash(
	\bs{1} \oslash \hat{\bs{v}}_{\Sf{z}}^{-, t}
	-
	\bs{1} \oslash \bs{v}_{\Sf{z}}^{+, t}
)
,\label{Eq:Update_v_s_z}\\
\bs{m}_{\Sf{z}}^{-, t}
= &
\bs{v}_{\Sf{z}}^{-, t} \odot(
	\hat{\bs{m}}_{\Sf{z}}^{-, t}
	\oslash
	\hat{\bs{v}}_{\Sf{z}}^{-, t}
	-
	\bs{m}_{\Sf{z}}^{+, t}
	\oslash
	\bs{v}_{\Sf{z}}^{+, t}
)
.\label{Eq:Update_m_s_z}
\end{align}

\TB{Step 2.2}: We can rewrite \eqref{Eq:P_s1_xi} as
\begin{align}
\bs{\Mbar}
\triangleq &
\check{\bs{x}}_{i}^{\Cts} \bs{A}_{i}
\sim
\NormalC[
	\bs{\Mbar} |
	(\bs{m}_{\check{\Sf{x}}, i}^{+, t})^{\Cts}
	\bs{A}_{i},
	\bs{A}_{i}^{\Cts}
	\bs{C}_{\check{\Sf{x}}, i}^{+, t}
	\bs{A}_{i}
]
,\label{Eq:cxi_Ai}\\
p_{\Sf{x}, i}^{-, 1, t}(\bs{x})
= &
\int{\dd z_{i} \dd \check{\bs{x}}_{i}}\,
\delta(
	z_{i}
	-
	\check{\bs{x}}_{i}^{\Cts} \bs{A}_{i} \bs{x}
)
p_{\Sf{z}}^{-, t}(z_{i})
p^{+, t}_{\check{\Sf{x}}}(\check{\bs{x}}_{i})
\nonumber\\
= &
\int{\dd \check{\bs{x}}_{i}}\,
\NormalC[
	\check{\bs{x}}_{i}^{\Cts} \bs{A}_{i} \bs{x} |
	\bs{m}_{\Sf{z}}^{-, t},
	\bs{v}_{\Sf{z}}^{-, t}
]
p^{+, t}_{\check{\Sf{x}}}(\check{\bs{x}}_{i})
\nonumber\\
\propto &
\int{\dd \bs{\Mbar}}\,
\NormalC[
	\bs{\Mbar} \bs{x} |
	\bs{m}_{\Sf{z}}^{-, t},
	\bs{v}_{\Sf{z}}^{-, t}
]
\NormalC[
	\bs{\Mbar} |
	(\bs{m}_{\check{\Sf{x}}, i}^{+, t})^{\Cts}
	\bs{A}_{i},
	\bs{A}_{i}^{\Cts}
	\bs{C}_{\check{\Sf{x}}, i}^{+, t}
	\bs{A}_{i}
]
\nonumber\\
\propto &
\EXP(
	-
	\frac{1}{2}
	\bs{x}^{\Cts}
	\bs{\Lambda}_{\Sf{x}, i}^{-, 1, t}
	\bs{x}
	+
	\bs{x}^{\Cts}
	\bs{b}_{\Sf{x}, i}^{-, 1, t}
)
,\label{Eq:Update_P_s1_xi}
\end{align}
where
\begin{align}
\bs{C}_{\check{\Sf{x}}, i}^{+, t}
= &
(\bs{\Lambda}_{\check{\Sf{x}}, i}^{+, t})^{-1}
,\label{Eq:Update_C_p_cxi}\\
\bs{m}_{\check{\Sf{x}}, i}^{+, t}
= &
\bs{C}_{\check{\Sf{x}}, i}^{+, t}
\bs{b}_{\check{\Sf{x}}, i}^{+, t}
,\label{Eq:Update_m_p_cxi}\\
a_{\check{\Sf{x}}, i}^{-, t}
= &
\frac{
	(\bs{m}_{\check{\Sf{x}}, i}^{+, t})^{\Cts}
	\bs{A}_{i}
	\bs{A}_{i}^{\Cts}
	\bs{C}_{\check{\Sf{x}}, i}^{+, t}
	\bs{A}_{i}
	\bs{A}_{i}^{\Cts}
	\bs{m}_{\check{\Sf{x}}, i}^{+, t}
}{
	|
		(\bs{m}_{\check{\Sf{x}}, i}^{+, t})^{\Cts}
		\bs{A}_{i}
		\bs{A}_{i}^{\Cts}
		\bs{m}_{\check{\Sf{x}}, i}^{+, t}
	|^{2}
}
,\label{Eq:Update_a_s_cxi}\\
\bs{b}_{\Sf{x}, i}^{-, 1, t}
= &
\frac{
	m_{\Sf{z}, i}^{-, t}
	\bs{A}_{i}^{\Cts}
	\bs{m}_{\check{\Sf{x}}, i}^{+, t}	
}{
	v_{\Sf{z}, i}^{-, t}
	+
	|m_{\Sf{z}, i}^{-, t}|^{2}
	a_{\check{\Sf{x}}, i}^{-, t}
}
,\label{Eq:Update_b_s1_xi}\\
\bs{\Lambda}_{\Sf{x}, i}^{-, 1, t}
= &
\frac{
	\bs{A}_{i}^{\Cts}
	\bs{m}_{\check{\Sf{x}}, i}^{+, t}
	(\bs{m}_{\check{\Sf{x}}, i}^{+, t})^{\Cts}
	\bs{A}_{i}
}{
	v_{\Sf{z}, i}^{-, t}
	+
	|m_{\Sf{z}, i}^{-, t}|^{2}
	a_{\check{\Sf{x}}, i}^{-, t}
}
.\label{Eq:Update_L_s1_xi}
\end{align}

\TB{Step 2.3}: Similarly, we can analyze \eqref{Eq:P_s_cx} as
\begin{align}
\bs{x}^{\Cts} \bs{A}_{i}^{\Cts}
\sim &
\NormalC[
	(\bs{m}_{\Sf{x}, i}^{+, 1, t})^{\Cts}
	\bs{A}_{i}^{\Cts},
	\bs{A}_{i}
	\bs{C}_{\Sf{x}, i}^{+, 1, t}
	\bs{A}_{i}^{\Cts}
]
,\label{Eq:xT_AiT}\\
p^{-, t}_{\check{\Sf{x}}}(\check{\bs{x}}_{i})
= &
\int{\dd z_{i} \dd \bs{x}}\,
\delta(
	z_{i}
	-
	\check{\bs{x}}_{i}^{\Cts} \bs{A}_{i} \bs{x}
)
p_{\Sf{z}}^{-, t}(z_{i})
p_{\Sf{x}, i}^{+, 1, t}(\bs{x})
\nonumber\\
= &
\int{\dd z_{i} \dd \bs{x}}\,
\delta(
	z_{i}^{\star}
	-
	\bs{x}^{\Cts} \bs{A}_{i}^{\Cts}
	\check{\bs{x}}_{i}
)
p_{\Sf{z}}^{-, t}(z_{i})
p_{\Sf{x}, i}^{+, 1, t}(\bs{x})
\nonumber\\
\propto &
\EXP(
	-
	\frac{1}{2}
	\check{\bs{x}}_{i}^{\Cts}
	\bs{\Lambda}_{\check{\Sf{x}}, i}^{-, t}
	\check{\bs{x}}_{i}
	+
	\check{\bs{x}}_{i}^{\Cts}
	\bs{b}_{\check{\Sf{x}}, i}^{-, t}
)
,\label{Eq:Update_P_s_cx}
\end{align}
where
\begin{align}
\bs{C}_{\Sf{x}, i}^{+, 1, t}
= &
(\bs{\Lambda}_{\Sf{x}, i}^{+, 1, t})^{-1}
,\label{Eq:Update_C_p1_xi}\\
\bs{m}_{\Sf{x}, i}^{+, 1, t}
= &
\bs{C}_{\Sf{x}, i}^{+, 1, t}
\bs{b}_{\Sf{x}, i}^{+, 1, t}
,\label{Eq:Update_m_p1_xi}\\
a_{\Sf{x}, i}^{-, 1, t}
=&
\frac{
	(\bs{m}_{\Sf{x}, i}^{+, 1, t})^{\Cts}
	\bs{A}_{i}^{\Cts}
	\bs{A}_{i}
	\bs{C}_{\Sf{x}, i}^{+, 1, t}
	\bs{A}_{i}^{\Cts}
	\bs{A}_{i}
	\bs{m}_{\Sf{x}, i}^{+, 1, t}
}{
	|
		(\bs{m}_{\Sf{x}, i}^{+, 1, t})^{\Cts}
		\bs{A}_{i}^{\Cts}
		\bs{A}_{i}
		\bs{m}_{\Sf{x}, i}^{+, 1, t}
	|^{2}
}
,\label{Eq:Update_a_s1_xi}\\
\bs{b}_{\check{\Sf{x}}, i}^{-, t}
= &
\frac{
	(m_{\Sf{z}, i}^{-, t})^{\star}
	\bs{A}_{i}
	\bs{m}_{\Sf{x}, i}^{+, 1, t}
}{
	v_{\Sf{z}, i}^{-, t}
	+
	|m_{\Sf{z}, i}^{-, t}|^{2}
	a_{\Sf{x}, i}^{-, 1, t}
}
,\label{Eq:Update_b_s_cxi}\\
\bs{\Lambda}_{\check{\Sf{x}}, i}^{-, t}
= &
\frac{
	\bs{A}_{i}
	\bs{m}_{\Sf{x}, i}^{+, 1, t}
	(\bs{m}_{\Sf{x}, i}^{+, 1, t})^{\Cts}
	\bs{A}_{i}^{\Cts}
}{
	v_{\Sf{z}, i}^{-, t}
	+
	|m_{\Sf{z}, i}^{-, t}|^{2}
	a_{\Sf{x}, i}^{-, 1, t}
}
.\label{Eq:Update_L_s_cxi}
\end{align}

\TB{Step 2.4}: We move to handle \eqref{Eq:P_s2_xi} as
\begin{align}
p_{\Sf{x}, i}^{-, 2, t}(\bs{x})
= &
\int{\dd \check{\bs{x}}_{i}}\,
\delta(\check{\bs{x}}_{i} - \bs{x})
p^{-, t}_{\check{\Sf{x}}}(\check{\bs{x}}_{i})
\propto
\EXP(
	-
	\frac{1}{2}
	\bs{x}^{\Cts}
	\bs{\Lambda}_{\Sf{x}, i}^{-, 2, t}
	\bs{x}
	+
	\bs{x}^{\Cts} \bs{b}_{\Sf{x}, i}^{-, 2, t}
)
,\label{Eq:Update_P_s2_xi}
\end{align}
where
\begin{align}
\bs{b}_{\Sf{x}, i}^{-, 2, t}
= &
\bs{b}_{\check{\Sf{x}}, i}^{-, t}
=
\frac{
	(m_{\Sf{z}, i}^{-, t})^{\star}
	\bs{A}_{i} \bs{m}_{\Sf{x}, i}^{+, 1, t}
}{
	v_{\Sf{z}, i}^{-, t}
	+
	|m_{\Sf{z}, i}^{-, t}|^{2}
	a_{\Sf{x}, i}^{-, 1, t}
}
,\label{Eq:Update_b_s2_xi}\\
\bs{\Lambda}_{\Sf{x}, i}^{-, 2, t}
= &
\bs{\Lambda}_{\check{\Sf{x}}, i}^{-, t}
=
\frac{
	\bs{A}_{i}
	\bs{m}_{\Sf{x}, i}^{+, 1, t}
	(\bs{m}_{\Sf{x}, i}^{+, 1, t})^{\Cts}
	\bs{A}_{i}^{\Cts}
}{
	v_{\Sf{z}, i}^{-, t}
	+
	|m_{\Sf{z}, i}^{-, t}|^{2}
	a_{\Sf{x}, i}^{-, 1, t}
}
.\label{Eq:Update_L_s2_xi}
\end{align}

\TB{Step 2.5}: Then substituting \eqref{Eq:Update_P_s1_xi} and \eqref{Eq:Update_P_s2_xi} into \eqref{Eq:P_s_x}, obtains
\begin{align}
p_{\Sf{x}}^{-, t}(\bs{x})
= &
\prod_{i = 1}^{M}{}[
	p_{\Sf{x}, i}^{-, 1, t}(\bs{x})
	p_{\Sf{x}, i}^{-, 2, t}(\bs{x})
]
\propto
\EXP(
	-
	\frac{1}{2}
	\bs{x}^{\Cts} \bs{\Lambda}_{\Sf{x}}^{-, t} \bs{x}
	+
	\bs{x}^{\Cts} \bs{b}_{\Sf{x}}^{-, t}
)
,\label{Eq:Update_P_s_x}
\end{align}
where
\begin{align}
\bs{b}_{\Sf{x}}^{-, t}
= &
\sum_{i = 1}^{M}{}(
	\bs{b}_{\Sf{x}, i}^{-, 1, t}
	+
	\bs{b}_{\Sf{x}, i}^{-, 2, t}
)
=
\sum_{i = 1}^{M}{}[
	\frac{
		m_{\Sf{z}, i}^{-, t}
		\bs{A}_{i}^{\Cts}
		\bs{m}_{\check{\Sf{x}}, i}^{+, t}
	}{
		v_{\Sf{z}, i}^{-, t}
		+
		|m_{\Sf{z}, i}^{-, t}|^{2}
		a_{\check{\Sf{x}}, i}^{-, t}
	}
	+
	\frac{
		(m_{\Sf{z}, i}^{-, t})^{\star}
		\bs{A}_{i} \bs{m}_{\Sf{x}, i}^{+, 1, t}
	}{
		v_{\Sf{z}, i}^{-, t}
		+
		|m_{\Sf{z}, i}^{-, t}|^{2}
		a_{\Sf{x}, i}^{-, 1, t}
	}
]
,\label{Eq:Update_b_s_x}\\
\bs{\Lambda}_{\Sf{x}}^{-, t}
= &
\sum_{i = 1}^{M}{}(
	\bs{\Lambda}_{\Sf{x}, i}^{-, 1, t}
	+
	\bs{\Lambda}_{\Sf{x}, i}^{-, 2, t}
)
=
\sum_{i = 1}^{M}{}[
	\frac{
		\bs{A}_{i}^{\Cts}
		\bs{m}_{\check{\Sf{x}}, i}^{+, t}
		(\bs{m}_{\check{\Sf{x}}, i}^{+, t})^{\Cts}
		\bs{A}_{i}
	}{
		v_{\Sf{z}, i}^{-, t}
		+
		|m_{\Sf{z}, i}^{-, t}|^{2}
		a_{\check{\Sf{x}}, i}^{-, t}
	}
	+
	\frac{
		\bs{A}_{i} \bs{m}_{\Sf{x}, i}^{+, 1, t}
		(\bs{m}_{\Sf{x}, i}^{+, 1, t})^{\Cts}
		\bs{A}_{i}^{\Cts}
	}{
		v_{\Sf{z}, i}^{-, t}
		+
		|m_{\Sf{z}, i}^{-, t}|^{2}
		a_{\Sf{x}, i}^{-, 1, t}
	}
]
.\label{Eq:Update_L_s_x}
\end{align}

\TB{Step 2.6}: We move on to handle \eqref{Eq:P_s_tx} and firstly compute $
f_{\tilde{\Sf{x}}}^{-, t}(\tilde{\bs{x}})
$ as
\begin{align}
f_{\tilde{\Sf{x}}}^{-, t}(\tilde{\bs{x}})
= &
\int{\dd \bs{x}}\,
\delta(\bs{x} - \tilde{\bs{x}})
p_{\Sf{x}}^{-, t}(\bs{x})
\propto
\EXP(
	-\frac{1}{2}
	\tilde{\bs{x}}^{\Cts}
	\bs{\Lambda}_{\tilde{\Sf{x}}}^{-, t}
	\tilde{\bs{x}}
	+
	\tilde{\bs{x}}^{\Cts}
	\bs{b}_{\tilde{\Sf{x}}}^{-, t}
)
,\label{Eq:Update_f_s_tx}
\end{align}
where
\begin{align}
\bs{b}_{\tilde{\Sf{x}}}^{-, t}
= &
\bs{b}_{\Sf{x}}^{-, t}
=
\sum_{i = 1}^{M}{}[
	\frac{
		m_{\Sf{z}, i}^{-, t}
		\bs{A}_{i}^{\Cts}
		\bs{m}_{\check{\Sf{x}}, i}^{+, t}	
	}{
		v_{\Sf{z}, i}^{-, t}
		+
		|m_{\Sf{z}, i}^{-, t}|^{2}
		a_{\check{\Sf{x}}, i}^{-, t}
	}
	+
	\frac{
		(m_{\Sf{z}, i}^{-, t})^{\star}
		\bs{A}_{i}
		\bs{m}_{\Sf{x}, i}^{+, 1, t}
	}{
		v_{\Sf{z}, i}^{-, t}
		+
		|m_{\Sf{z}, i}^{-, t}|^{2}
		a_{\Sf{x}, i}^{-, 1, t}
	}
]
,\label{Eq:Update_b_s_tx}\\
\bs{\Lambda}_{\tilde{\Sf{x}}}^{-, t}
= &
\bs{\Lambda}_{\Sf{x}}^{-, t}
=
\sum_{i = 1}^{M}{}[
	\frac{
		\bs{A}_{i}^{\Cts}
		\bs{m}_{\check{\Sf{x}}, i}^{+, t}
		(\bs{m}_{\check{\Sf{x}}, i}^{+, t})^{\Cts}
		\bs{A}_{i}
	}{
		v_{\Sf{z}, i}^{-, t}
		+
		|m_{\Sf{z}, i}^{-, t}|^{2}
		a_{\check{\Sf{x}}, i}^{-, t}
	}
	+
	\frac{
		\bs{A}_{i}
		\bs{m}_{\Sf{x}, i}^{+, 1, t}
		(\bs{m}_{\Sf{x}, i}^{+, 1, t})^{\Cts}
		\bs{A}_{i}^{\Cts}
	}{
		v_{\Sf{z}, i}^{-, t}
		+
		|m_{\Sf{z}, i}^{-, t}|^{2}
		a_{\Sf{x}, i}^{-, 1, t}
	}
]
,\label{Eq:Update_L_s_tx}
\end{align}
then the projection function in the numerator of \eqref{Eq:P_s_tx} denotes as $
\NormalC[
	\tilde{\bs{x}} |
	\hat{\bs{m}}_{\tilde{\Sf{x}}}^{-, t},
	\hat{\bs{v}}_{\tilde{\Sf{x}}}^{-, t}
]
$ with
\begin{align}
\hat{\bs{C}}_{\tilde{\Sf{x}}}^{-, t}
= &
[
	\bs{\Lambda}_{\tilde{\Sf{x}}}^{-, t}
	+
	\Diag(
		\bs{1} \oslash \bs{v}_{\tilde{\Sf{x}}}^{+, t}
	)
]^{-1}
,\label{Eq:Update_HC_s_tx}\\
\hat{\bs{m}}_{\tilde{\Sf{x}}}^{-, t}
= &
\hat{\bs{C}}_{\tilde{\Sf{x}}}^{-, t}
(
	\bs{b}_{\tilde{\Sf{x}}}^{-, t}
	+
	\bs{m}_{\tilde{\Sf{x}}}^{+, t}
	\oslash
	\bs{v}_{\tilde{\Sf{x}}}^{+, t}
)
,\label{Eq:Update_Hm_s_tx}\\
\hat{\bs{v}}_{\tilde{\Sf{x}}}^{-, t}
=&
\diag(
	\hat{\bs{C}}_{\tilde{\Sf{x}}}^{-, t}
)
,\label{Eq:Update_Hv_s_tx}
\end{align}
and the entirety of \eqref{Eq:P_s_tx} is computed via the Gaussian reproduction property as
\begin{align}
p_{\tilde{\Sf{x}}}^{-, t}(\tilde{\bs{x}})
= &
\NormalC[
	\tilde{\bs{x}} |
		\bs{m}_{\tilde{\Sf{x}}}^{-, t},
		\bs{v}_{\tilde{\Sf{x}}}^{-, t}
]
,\label{Eq:Update_P_s_tx}
\end{align}
where
\begin{align}
\bs{v}_{\tilde{\Sf{x}}}^{-, t}
= &
\bs{1} \oslash
(
	\bs{1} \oslash
	\hat{\bs{v}}_{\tilde{\Sf{x}}}^{-, t}
	-
	\bs{1} \oslash
	\bs{v}_{\tilde{\Sf{x}}}^{+, t}
)
,\label{Eq:Update_v_s_tx}\\
\bs{m}_{\tilde{\Sf{x}}}^{-, t}
= &
\bs{v}_{\tilde{\Sf{x}}}^{-, t}
\odot
(
	\hat{\bs{m}}_{\tilde{\Sf{x}}}^{-, t}
	\oslash
	\hat{\bs{v}}_{\tilde{\Sf{x}}}^{-, t}
	-
	\bs{m}_{\tilde{\Sf{x}}}^{+, t}
	\oslash
	\bs{v}_{\tilde{\Sf{x}}}^{+, t}
).\label{Eq:Update_m_s_tx}
\end{align}

For \TB{S3}, we present the procedure of deriving all forward passing messages as follows.

\TB{Step 3.1}: We denote the projection function in the numerator of \eqref{Eq:P_p_tx} as $
\NormalC[
	\tilde{\bs{x}} |
	\hat{\bs{m}}_{\tilde{\Sf{x}}}^{+, t + 1},
	\hat{\bs{v}}_{\tilde{\Sf{x}}}^{+, t + 1}
]
$, where $
\hat{\bs{m}}_{\tilde{\Sf{x}}}^{+, t + 1}
=
\Mean[
	\tilde{\bs{x}} |
	\bs{m}_{\tilde{\Sf{x}}}^{-, t},
	\bs{v}_{\tilde{\Sf{x}}}^{-, t}
]
$ and $
\hat{\bs{v}}_{\tilde{\Sf{x}}}^{+, t + 1}
=
\Var[
	\tilde{\bs{x}} |
	\bs{m}_{\tilde{\Sf{x}}}^{-, t},
	\bs{v}_{\tilde{\Sf{x}}}^{-, t}
]
$ are the mean and variance of the approximate posterior distribution $
p(\tilde{\bs{x}} | \bs{y})
=
\frac{
	p(\tilde{\bs{x}})
	\NormalC[
		\tilde{\bs{x}} |
		\bs{m}_{\tilde{\Sf{x}}}^{-, t},
		\bs{v}_{\tilde{\Sf{x}}}^{-, t}
	]
}{
	\int{\dd \tilde{\bs{x}}}\,
	p(\tilde{\bs{x}})
	\NormalC[
		\tilde{\bs{x}} |
		\bs{m}_{\tilde{\Sf{x}}}^{-, t},
		\bs{v}_{\tilde{\Sf{x}}}^{-, t}
	]
}
$, respectively. Then the entirety of \eqref{Eq:P_p_tx} is computed via the Gaussian reproduction property as
\begin{align}
p_{\tilde{\Sf{x}}}^{+, t + 1}(\tilde{\Sf{x}})
= &
\NormalC[
	\tilde{\bs{x}} |
	\bs{m}_{\tilde{\Sf{x}}}^{+, t + 1},
	\bs{v}_{\tilde{\Sf{x}}}^{+, t + 1}
],
\label{Eq:Update_p_p_tx}
\end{align}
where
\begin{align}
\bs{v}_{\tilde{\Sf{x}}}^{+, t + 1}
= &
\bs{1} \oslash
(
	\bs{1} \oslash
	\hat{\bs{v}}_{\tilde{\Sf{x}}}^{+, t + 1}
	-
	\bs{1} \oslash
	\bs{v}_{\tilde{\Sf{x}}}^{-, t}
)
,\label{Eq:Update_v_p_tx}\\
\bs{m}_{\tilde{\Sf{x}}}^{+, t + 1}
= &
\bs{v}_{\tilde{\Sf{x}}}^{+, t + 1} \odot
(
	\hat{\bs{m}}_{\tilde{\Sf{x}}}^{+, t + 1}
	\oslash
	\hat{\bs{v}}_{\tilde{\Sf{x}}}^{+, t + 1}
	-
	\bs{m}_{\tilde{\Sf{x}}}^{-, t}
	\oslash
	\bs{v}_{\tilde{\Sf{x}}}^{-, t}
)
.\label{Eq:Update_m_p_tx}
\end{align}

\TB{Step 3.2}: We continue to analyze \eqref{Eq:P_p_x} as
\begin{align}
p_{\Sf{x}}^{+, t + 1}(\bs{x})
= &
\int{\dd \tilde{\bs{x}}}\,
\delta(\bs{x} - \tilde{\bs{x}})
p_{\tilde{\Sf{x}}}^{+, t + 1}(\tilde{\bs{x}})
=
\NormalC[
	\bs{x} |
	\bs{m}_{\Sf{x}}^{+, t + 1},
	\bs{v}_{\Sf{x}}^{+, t + 1}
]
,\label{Eq:Update_P_p_x}
\end{align}
where
\begin{align}
\bs{m}_{\Sf{x}}^{+, t + 1}
= &
\bs{m}_{\tilde{\Sf{x}}}^{+, t + 1}
,\label{Eq:Update_m_p_x}\\
\bs{v}_{\Sf{x}}^{+, t + 1}
= &
\bs{v}_{\tilde{\Sf{x}}}^{+, t + 1}
.\label{Eq:Update_v_p_x}
\end{align}

\TB{Step 3.3}: Then we handle \eqref{Eq:P_p1_xi} as
\begin{align}
p_{\Sf{x},i}^{+, 1, t + 1}(\bs{x})
= &
p_{\Sf{x}}^{+, t + 1}(\bs{x})
[
	\prod_{k \neq i}^{M}{}
	p_{\Sf{x}, k}^{-, 1, t}(\bs{x})
]
[
	\prod_{k = 1}^{M}{}
	p_{\Sf{x}, k}^{-, 2, t}(\bs{x})
]
\nonumber\\
\propto &
p_{\Sf{x}}^{+, t + 1}(\bs{x})
\EXP(
	-
	\frac{1}{2}
	\bs{x}^{\Cts}
	\bs{\Lambda}_{\Sf{x} \Bs i}^{-, 1, t}
	\bs{x}
	+
	\bs{x}^{\Cts} \bs{b}_{\Sf{x} \Bs i}^{-, 1, t}
)
\nonumber\\
= &
\NormalC[
	\bs{x} |
	\bs{m}_{\Sf{x}, i}^{+, 1, t + 1},
	\bs{C}_{\Sf{x}, i}^{+, 1, t + 1}
]
,\label{Eq:Update_P_p1_xi}
\end{align}
where
\begin{align}
\bs{b}_{\Sf{x} \Bs i}^{-, 1, t}
= &
\sum_{k \neq i}^{M}{}
\bs{b}_{\Sf{x}, k}^{-, 1, t}
+
\sum_{k = 1}^{M}{}
\bs{b}_{\Sf{x}, k}^{-, 2, t}
=
\sum_{k \neq i}^{M}{}
\frac{
	m_{\Sf{z}, k}^{-, t}
	\bs{A}_{k}^{\Cts}
	\bs{m}_{\check{\Sf{x}}, k}^{+, t}
}{
	v_{\Sf{z}, k}^{-, t}
	+
	|m_{\Sf{z}, k}^{-, t}|^{2}
	a_{\check{\Sf{x}}, k}^{-, t}
}
+
\sum_{k = 1}^{M}{}
\frac{
	(m_{\Sf{z}, k}^{-, t})^{\star}
	\bs{A}_{k} \bs{m}_{\Sf{x}, k}^{+, 1, t}
}{
	v_{\Sf{z}, k}^{-, t}
	+
	|m_{\Sf{z}, k}^{-, t}|^{2}
	a_{\Sf{x}, k}^{-, 1, t}
}
,\label{Eq:Update_b_s1_xsi}\\
\bs{\Lambda}_{\Sf{x} \Bs i}^{-, 1, t}
= &
\sum_{k \neq i}^{M}{}
\bs{\Lambda}_{\Sf{x}, k}^{-, 1, t}
+
\sum_{k =1}^{M}{}
\bs{\Lambda}_{\Sf{x}, k}^{-, 2, t}
=
\sum_{k \neq i}^{M}{}
\frac{
	\bs{A}_{k}^{\Cts}
	\bs{m}_{\check{\Sf{x}}, k}^{+, t}
	(\bs{m}_{\check{\Sf{x}}, k}^{+, t})^{\Cts}
	\bs{A}_{k}
}{
	v_{\Sf{z}, k}^{-, t}
	+
	|m_{\Sf{z}, k}^{-, t}|^{2}
	a_{\check{\Sf{x}}, k}^{-, t}
}
+
\sum_{k = 1}^{M}{}
\frac{
	\bs{A}_{k} \bs{m}_{\Sf{x}, k}^{+, 1, t}
	(\bs{m}_{\Sf{x}, k}^{+, 1, t})^{\Cts}
	\bs{A}_{k}^{\Cts}
}{
	v_{\Sf{z}, k}^{-, t}
	+
	|m_{\Sf{z}, k}^{-, t}|^{2}
	a_{\Sf{x}, k}^{-, 1, t}
}
,\label{Eq:Update_L_s1_xsi}\\
\bs{b}_{\Sf{x}, i}^{+, 1, t + 1}
= &
\bs{b}_{\Sf{x} \Bs i}^{-, 1, t}
+
\bs{m}_{\Sf{x}}^{+, t + 1}
\oslash
\bs{v}_{\Sf{x}}^{+, t + 1}
,\label{Eq:Update_b_p1_xi}\\
\bs{C}_{\Sf{x}, i}^{+, 1, t + 1}
= &
[
	\bs{\Lambda}_{\Sf{x} \Bs i}^{-, 1, t}
	+
	\Diag(\bs{1} \oslash \bs{v}_{\Sf{x}}^{+, t + 1})
]^{- 1}
,\label{Eq:Update_C_p1t1_xi}\\
\bs{m}_{\Sf{x}, i}^{+, 1, t + 1}
= &
\bs{C}_{\Sf{x}, i}^{+, 1, t + 1}
\bs{b}_{\Sf{x}, i}^{+, 1, t + 1}
.\label{Eq:Update_m_p1t1_xi}
\end{align}

For simplifying \eqref{Eq:Update_P_p1_xi}, the corresponding broadcast versions of \eqref{Eq:Update_b_p1_xi}-\eqref{Eq:Update_m_p1t1_xi} denote as 
\begin{align}
\hat{\bs{b}}_{\Sf{x}}^{+, t + 1}
= &
\bs{b}_{\Sf{x}}^{-, t}
+
\bs{m}_{\Sf{x}}^{+, t + 1}
\oslash
\bs{v}_{\Sf{x}}^{+, t + 1}
,\label{Eq:Update_Hb_p_x}\\
\hat{\bs{C}}_{\Sf{x}}^{+, t + 1}
= &
[
	\bs{\Lambda}_{\Sf{x}}^{-, t}
	+
	\Diag(\bs{1} \oslash \bs{v}_{\Sf{x}}^{+, t + 1})
]^{- 1}
,\label{Eq:Update_HC_p_x}\\
\hat{\bs{m}}_{\Sf{x}}^{+, t + 1}
= &
\hat{\bs{C}}_{\Sf{x}}^{+, t + 1}
\hat{\bs{b}}_{\Sf{x}}^{+, t + 1}
.\label{Eq:Update_Hm_p_x}
\end{align}

Substituting \eqref{Eq:Update_Hb_p_x}-\eqref{Eq:Update_Hm_p_x} into \eqref{Eq:Update_C_p1t1_xi}, yields
\begin{align}
\bs{C}_{\Sf{x}, i}^{+, 1, t + 1}
= &
[
	\bs{\Lambda}_{\Sf{x} \Bs i}^{-, 1, t}
	+
	\Diag(\bs{1} \oslash \bs{v}_{\Sf{x}}^{+, t + 1})
]^{- 1}
\nonumber\\
= &
[
	\bs{\Lambda}_{\Sf{x}}^{-, 1, t}
	+
	\Diag(\bs{1} \oslash \bs{v}_{\Sf{x}}^{+, t + 1})
	-
	\bs{\Lambda}_{\Sf{x}, i}^{-, 1, t}
]^{- 1}
\nonumber\\
= &
[
	(
		\hat{\bs{C}}_{\Sf{x}}^{+, t + 1}
	)^{- 1}
	-
	\bs{\Lambda}_{\Sf{x}, i}^{-, 1, t}
]^{- 1}
\nonumber\\
= &
[
	(
		\hat{\bs{C}}_{\Sf{x}}^{+, t + 1}
	)^{- 1}
	-
	\frac{
		\bs{A}_{i}^{\Cts}
		\bs{m}_{\check{\Sf{x}}, i}^{+, t}
		(\bs{m}_{\check{\Sf{x}}, i}^{+, t})^{\Cts}
		\bs{A}_{i}
	}{
		v_{\Sf{z}, i}^{-, t}
		+
		|m_{\Sf{z}, i}^{-, t}|^{2}
		a_{\check{\Sf{x}}, i}^{-, t}
	}
]^{- 1}
\nonumber\\
\overset{\text{(A)}}{=} &
\hat{\bs{C}}_{\Sf{x}}^{+, t + 1}
+
\frac{
	\hat{\bs{C}}_{\Sf{x}}^{+, t + 1}
	\bs{A}_{i}^{\Cts}
	\bs{m}_{\check{\Sf{x}}, i}^{+, t}
	(\bs{m}_{\check{\Sf{x}}, i}^{+, t})^{\Cts}
	\bs{A}_{i}	
	\hat{\bs{C}}_{\Sf{x}}^{+, t + 1}
}{
	v_{\Sf{z}, i}^{-, t}
	+
	|m_{\Sf{z}, i}^{-, t}|^{2}
	a_{\check{\Sf{x}}, i}^{-, t}
	-
	(\bs{m}_{\check{\Sf{x}}, i}^{+, t})^{\Cts}
	\bs{A}_{i}
	\hat{\bs{C}}_{\Sf{x}}^{+, t + 1}
	\bs{A}_{i}^{\Cts}
	\bs{m}_{\check{\Sf{x}}, i}^{+, t}
}
,\label{Eq:Update_C_p1t1_xi_Reformula}
\end{align}
where (A) follows the Sherman-Morrison formulation.

Then substituting \eqref{Eq:Update_Hb_p_x} and \eqref{Eq:Update_C_p1t1_xi_Reformula} into \eqref{Eq:Update_m_p1t1_xi}, yields
\begin{align}
\bs{m}_{\Sf{x}, i}^{+, 1, t + 1}
= &
\bs{C}_{\Sf{x}, i}^{+, 1, t + 1}
\bs{b}_{\Sf{x}, i}^{+, 1, t + 1}
\nonumber\\
= &
[
	\hat{\bs{C}}_{\Sf{x}}^{+, t + 1}
	+
	\frac{
		\hat{\bs{C}}_{\Sf{x}}^{+, t + 1}
		\bs{A}_{i}^{\Cts}
		\bs{m}_{\check{\Sf{x}}, i}^{+, t}
		(\bs{m}_{\check{\Sf{x}}, i}^{+, t})^{\Cts}
		\bs{A}_{i}	
		\hat{\bs{C}}_{\Sf{x}}^{+, t + 1}
	}{
		v_{\Sf{z}, i}^{-, t}
		+
		|m_{\Sf{z}, i}^{-, t}|^{2}
		a_{\check{\Sf{x}}, i}^{-, t}
		-
		(\bs{m}_{\check{\Sf{x}}, i}^{+, t})^{\Cts}
		\bs{A}_{i}
		\hat{\bs{C}}_{\Sf{x}}^{+, t + 1}
		\bs{A}_{i}^{\Cts}
		\bs{m}_{\check{\Sf{x}}, i}^{+, t}
	}
]
(
	\hat{\bs{b}}_{\Sf{x}}^{+, t + 1}
	-
	\frac{
		m_{\Sf{z}, i}^{-, t}
		\bs{A}_{i}^{\Cts}
		\bs{m}_{\check{\Sf{x}}, i}^{+, t}
	}{
		v_{\Sf{z}, i}^{-, t}
		+
		|m_{\Sf{z}, i}^{-, t}|^{2}
		a_{\check{\Sf{x}}, i}^{-, t}
	}
)
\nonumber\\
= &
\hat{\bs{m}}_{\Sf{x}}^{+, t + 1}
-
\frac{
	m^{-, t}_{\Sf{z}, i}
	-
	(\bs{m}_{\check{\Sf{x}}, i}^{+, t})^{\Cts}
	\bs{A}_{i}
	\hat{\bs{m}}^{+, t + 1}_{\Sf{x}}
}{
	v_{\Sf{z}, i}^{-, t}
	+
	|m_{\Sf{z}, i}^{-, t}|^{2}
	a_{\check{\Sf{x}}, i}^{-, t}
	-
	(\bs{m}_{\check{\Sf{x}}, i}^{+, t})^{\Cts}
	\bs{A}_{i}
	\hat{\bs{C}}_{\Sf{x}}^{+, t + 1}
	\bs{A}_{i}^{\Cts}
	\bs{m}_{\check{\Sf{x}}, i}^{+, t}
}
\hat{\bs{C}}_{\Sf{x}}^{+, t + 1}
\bs{A}_{i}^{\Cts}\bs{m}_{\check{\Sf{x}}, i}^{+, t}
.\label{Eq:Update_m_p1t1_xi_Reformula}
\end{align}

\TB{Step 3.4}: We similarly denote \eqref{Eq:P_p2_xi} as
\begin{align}
p_{\Sf{x}, i}^{+, 2, t + 1}(\bs{x})
= &
p_{\Sf{x}}^{+, t + 1}(\bs{x})
[
	\prod_{k = 1}^{M}{}
	p_{\Sf{x}, k}^{-, 1, t}(\bs{x})
]
[
	\prod_{k \neq i}^{M}{}
	p_{\Sf{x}, k}^{-, 2, t}(\bs{x})
]
=
\NormalC[
	\bs{x} |
	\bs{m}_{\Sf{x}, i}^{+, 2, t + 1},
	\bs{C}_{\Sf{x}, i}^{+, 2, t + 1}
]
,\label{Eq:Update_P_p2_xi}
\end{align}
where
\begin{align}
\bs{C}_{\Sf{x}, i}^{+, 2, t + 1}
= &
\hat{\bs{C}}_{\Sf{x}}^{+, t + 1}
+
\frac{
	\hat{\bs{C}}_{\Sf{x}}^{+, t + 1}
	\bs{A}_{i} \bs{m}_{\Sf{x}, i}^{+, 1, t}
	(\bs{m}_{\Sf{x}, i}^{+, 1, t})^{\Cts}
	\bs{A}_{i}^{\Cts}	
	\hat{\bs{C}}_{\Sf{x}}^{+, t + 1}
}{
	v_{\Sf{z}, i}^{-, t}
	+
	|m_{\Sf{z}, i}^{-, t}|^{2}
	a_{\Sf{x}, i}^{-, 1, t}
	-
	(\bs{m}_{\Sf{x}, i}^{+, 1, t})^{\Cts}
	\bs{A}_{i}^{\Cts}
	\hat{\bs{C}}_{\Sf{x}}^{+, t + 1}
	\bs{A}_{i} \bs{m}_{\Sf{x}, i}^{+, 1, t}
}
,\label{Eq:Update_C_p2_xi}\\
\bs{m}_{\Sf{x}, i}^{+, 2, t + 1}
= &
\hat{\bs{m}}_{\Sf{x}}^{+, t + 1}
-
\frac{
	(m^{-, t}_{\Sf{z}, i})^{\star}
	-
	(\bs{m}_{\Sf{x}, i}^{+, 1, t})^{\Cts}
	\bs{A}_{i}^{\Cts}
	\hat{\bs{m}}_{\Sf{x}}^{+, t + 1}
}{
	v_{\Sf{z}, i}^{-, t}
	+
	|m_{\Sf{z}, i}^{-, t}|^{2}
	a_{\Sf{x}, i}^{-, 1, t}
	-
	(\bs{m}_{\Sf{x}, i}^{+, 1, t})^{\Cts}
	\bs{A}_{i}^{\Cts}
	\hat{\bs{C}}_{\Sf{x}}^{+, t + 1}
	\bs{A}_{i} \bs{m}_{\Sf{x}, i}^{+, 1, t}
}
\hat{\bs{C}}^{+, t + 1}_{\Sf{x}}
\bs{A}_{i} \bs{m}_{\Sf{x}, i}^{+, 1, t}
.\label{Eq:Update_m_p2_xi}
\end{align}

\TB{Step 3.5}: We move to handle \eqref{Eq:P_p_cx} as
\begin{align}
p_{\check{\Sf{x}}}^{+, t + 1}(\check{\bs{x}}_{i})
= &
\int{\dd \bs{x}}\,
\delta(\check{\bs{x}}_{i} - \bs{x})
p_{\Sf{x}, i}^{+, 2, t + 1}(\bs{x})
=
\NormalC[
	\check{\bs{x}}_{i} |
	\bs{m}_{\check{\Sf{x}}, i}^{+, t + 1},
	\bs{C}_{\check{\Sf{x}}, i}^{+, t + 1}
]
,\label{Eq:Update_P_p_cx}
\end{align}
where 
\begin{align}
\bs{m}_{\check{\Sf{x}}, i}^{+, t + 1}
= &
\bs{m}_{\Sf{x}, i}^{+, 2, t + 1}
=
\hat{\bs{m}}_{\Sf{x}}^{+, t + 1}
-
\frac{
	(m^{-, t}_{\Sf{z}, i})^{\star}
	-
	(\bs{m}_{\Sf{x}, i}^{+, 1, t})^{\Cts}
	\bs{A}_{i}^{\Cts}
	\hat{\bs{m}}_{\Sf{x}}^{+, t + 1}
}{
	v_{\Sf{z}, i}^{-, t}
	+
	|m_{\Sf{z}, i}^{-, t}|^{2}
	a_{\Sf{x}, i}^{-, 1, t}
	-
	(\bs{m}_{\Sf{x}, i}^{+, 1, t})^{\Cts}
	\bs{A}_{i}^{\Cts}
	\hat{\bs{C}}_{\Sf{x}}^{+, t + 1}
	\bs{A}_{i}\bs{m}_{\Sf{x},i}^{+,1,t}
}
\hat{\bs{C}}^{+, t + 1}_{\Sf{x}}
\bs{A}_{i} \bs{m}_{\Sf{x}, i}^{+, 1, t}
,\label{Eq:Update_m_pt1_cxi}\\
\bs{C}_{\check{\Sf{x}}, i}^{+, t + 1}
= &
\bs{C}_{\Sf{x}, i}^{+, 2, t + 1}
=
\hat{\bs{C}}_{\Sf{x}}^{+, t + 1}
+
\frac{
	\hat{\bs{C}}_{\Sf{x}}^{+, t + 1}
	\bs{A}_{i} \bs{m}_{\Sf{x}, i}^{+, 1, t}
	(\bs{m}_{\Sf{x}, i}^{+, 1, t})^{\Cts}
	\bs{A}_{i}^{\Cts}	
	\hat{\bs{C}}_{\Sf{x}}^{+, t + 1}
}{
	v_{\Sf{z}, i}^{-, t}
	+
	|m_{\Sf{z}, i}^{-, t}|^{2}
	a_{\Sf{x}, i}^{-, 1, t}
	-
	(\bs{m}_{\Sf{x}, i}^{+, 1, t})^{\Cts}
	\bs{A}_{i}^{\Cts}
	\hat{\bs{C}}_{\Sf{x}}^{+, t + 1}
	\bs{A}_{i}\bs{m}_{\Sf{x},i}^{+,1,t}
}
.\label{Eq:Update_C_pt1_cxi}
\end{align}

\TB{Step 3.6}: Finally we compute \eqref{Eq:P_p_z} as
\begin{align}
p_{\Sf{z}}^{+, t + 1}(z_{i})
= &
\int{\dd \check{\bs{x}}_{i} \dd \bs{x}}\,
\delta(
	z_{i}
	-
	\check{\bs{x}}_{i}^{\Cts} \bs{A}_{i} \bs{x}
)
p_{\check{\Sf{x}}}^{+, t + 1}(\check{\bs{x}}_{i})
p_{\Sf{x}, i}^{+, 1, t + 1}(\bs{x})
=
\NormalC[
	z_{i} |
		m_{\Sf{z}}^{+, t + 1},
		v_{\Sf{z}}^{+, t + 1}
]
,\label{Eq:Update_P_p_z}
\end{align}
where
\begin{align}
m_{\Sf{z}, i}^{+, t + 1}
= &
\int{\dd z_{i}}\,
z_{i} p_{\Sf{z}}^{+, t + 1}(z_{i})
\nonumber\\
= &
\int{
	\dd z_{i}
	\dd \check{\bs{x}}_{i} \dd \bs{x}
}\,
z_{i}
\delta(
	z_{i}
	-
	\check{\bs{x}}_{i}^{\Cts} \bs{A}_{i} \bs{x}
)
p_{\check{\Sf{x}}}^{+, t + 1}(\check{\bs{x}}_{i})
p_{\Sf{x}, i}^{+, 1, t + 1}(\bs{x})
\nonumber\\
= &
\int{
	\dd \check{\bs{x}}_{i} \dd \bs{x}
}\,
\check{\bs{x}}_{i}^{\Cts} \bs{A}_{i} \bs{x}
p_{\check{\Sf{x}}}^{+, t + 1}(\check{\bs{x}}_{i})
p_{\Sf{x}, i}^{+, 1, t + 1}(\bs{x})
\nonumber\\
= &
(\bs{m}_{\check{\Sf{x}}, i}^{+, t + 1})^{\Cts}
\bs{A}_{i}
\bs{m}_{\Sf{x}, i}^{+, 1, t + 1}
,\label{Eq:Update_m_p_zi}\\
v_{\Sf{z}, i}^{+, t + 1}
= &
\int{\dd z_{i}}\,
z_{i}^{2}
p_{\Sf{z}}^{+, t + 1}(z_{i})
-
(m_{\Sf{z}, i}^{+, t + 1})^{2}
\nonumber\\
= &
\int{
	\dd z_{i}
	\dd \check{\bs{x}}_{i} \dd \bs{x}
}\,
z_{i}^{2}
\delta(
	z_{i}
	-
	\check{\bs{x}}_{i}^{\Cts} \bs{A}_{i} \bs{x}
)
p_{\check{\Sf{x}}}^{+, t + 1}(\check{\bs{x}}_{i})
p_{\Sf{x}, i}^{+, 1, t + 1}(\bs{x})
-
(m_{\Sf{z}, i}^{+, t + 1})^{2}
\nonumber\\
= &
\int{
	\dd \check{\bs{x}}_{i} \dd \bs{x}
}\,
\check{\bs{x}}_{i}^{\Cts} \bs{A}_{i} \bs{x}
\bs{x}^{\Cts} \bs{A}_{i}^{\Cts} \check{\bs{x}}_{i}
p_{\check{\Sf{x}}}^{+, t + 1}(\check{\bs{x}}_{i})
p_{\Sf{x}, i}^{+, 1, t + 1}(\bs{x})
-
(m_{\Sf{z}, i}^{+, t + 1})^{2}
\nonumber\\
= &
\int{
	\dd \check{\bs{x}}_{i} \dd \bs{x}
}\,
\Trace(
\bs{A}_{i}^{\Cts}
\check{\bs{x}}_{i} \check{\bs{x}}_{i}^{\Cts}
\bs{A}_{i}
\bs{x} \bs{x}^{\Cts}
)
p_{\check{\Sf{x}}}^{+, t + 1}(\check{\bs{x}}_{i})
p_{\Sf{x}, i}^{+, 1, t + 1}(\bs{x})
-
(m_{\Sf{z}, i}^{+, t + 1})^{2}
\nonumber\\
=&
\Trace(
	\bs{A}_{i}^{\Cts}
	\bs{C}_{\check{\Sf{x}}, i}^{+, t + 1}
	\bs{A}_{i}
	\bs{C}_{\Sf{x}, i}^{+,1, t + 1}
)
+
(\bs{m}_{\check{\Sf{x}}, i}^{+, t + 1})^{\Cts}
\bs{A}_{i}
\bs{C}_{\Sf{x}, i}^{+, 1, t + 1}
\bs{A}_{i}^{\Cts}
\bs{m}_{\check{\Sf{x}}, i}^{+, t + 1}
+
(\bs{m}_{\Sf{x}, i}^{+, 1, t + 1})^{\Cts}
\bs{A}_{i}^{\Cts}
\bs{C}_{\check{\Sf{x}}, i}^{+, t + 1}
\bs{A}_{i}
\bs{m}_{\Sf{x}, i}^{+, 1, t + 1}
.\label{Eq:Update_v_p_zi}
\end{align}

\begin{breakablealgorithm}
\setstretch{2}
\caption{The QMP algorithm}
\label{Algo:QMP}
\begin{algorithmic}[1]
\State
Input:
$\bs{y}$,
$\{\bs{A}_{i}\}_{i = 1, \cdots, M}$,
$p(\bs{y} | \bs{z})$\\
Initialize: $
\bs{m}_{\Sf{z}}^{+, 1}
$, $
\bs{v}_{\Sf{z}}^{+, 1}
$, $
\bs{m}_{\tilde{\Sf{x}}}^{+, 1}
$, $
\bs{v}_{\tilde{\Sf{x}}}^{+, 1}
$, $
\bs{m}_{\Sf{x}, i}^{+, 1, 1}
$, $
\bs{C}_{\Sf{x}, i}^{+, 1, 1}
$, $
\bs{m}_{\Sf{x}, i}^{+, 2, 1}
$, $
\bs{C}_{\Sf{x}, i}^{+, 2, 1}
$.\\
Iterate: $t = 1, \cdots, T$\\
$
(
	\hat{\bs{m}}_{\Sf{z}}^{-, t},
	\hat{\bs{v}}_{\Sf{z}}^{-, t}
)
=
\Mean[
	\bs{z} |
	\bs{m}_{\Sf{z}}^{+, t},
	\bs{v}_{\Sf{z}}^{+, t}
]
$\\
$
(
	\bs{m}_{\Sf{z}}^{-, t},
	\bs{v}_{\Sf{z}}^{-, t}
)
=
\Ext[
	\hat{\bs{m}}_{\Sf{z}}^{-, t},
	\hat{\bs{v}}_{\Sf{z}}^{-, t},
	\bs{m}_{\Sf{z}}^{+, t},
	\bs{v}_{\Sf{z}}^{+, t}
]
$\\
$
a_{\Sf{x}, i}^{-, 1, t}
=
\frac{
	(\bs{m}_{\Sf{x}, i}^{+, 1, t})^{\Cts}
	\bs{A}_{i}^{\Cts}
	\bs{A}_{i}
	\bs{C}_{\Sf{x}, i}^{+, 1, t}
	\bs{A}_{i}^{\Cts}
	\bs{A}_{i} \bs{m}_{\Sf{x}, i}^{+, 1, t}
}{
	|
		(\bs{m}_{\Sf{x}, i}^{+, 1, t})^{\Cts}
		\bs{A}_{i}^{\Cts}
		\bs{A}_{i} \bs{m}_{\Sf{x}, i}^{+, 1, t}
	|^{2}
}
$\\
$
a_{\Sf{x}, i}^{-, 2, t}
=
\frac{
	(\bs{m}_{\Sf{x}, i}^{+, 2, t})^{\Cts} \bs{A}_{i}
	\bs{A}_{i}^{\Cts}
	\bs{C}_{\Sf{x}, i}^{+, 2, t} \bs{A}_{i}
	\bs{A}_{i}^{\Cts} \bs{m}_{\Sf{x}, i}^{+, 2, t}
}{
	|
		(\bs{m}_{\Sf{x}, i}^{+, 2, t})^{\Cts}
		\bs{A}_{i}
		\bs{A}_{i}^{\Cts} \bs{m}_{\Sf{x}, i}^{+, 2, t}
	|^{2}
}
$\\
$
\bs{b}_{\tilde{\Sf{x}}}^{-, t}
=
\sum_{i = 1}^{M}{}[
	\frac{
		m_{\Sf{z}, i}^{-, t}
		\bs{A}_{i}^{\Cts}
		\bs{m}_{\Sf{x},i}^{+, 2, t}	
	}{
		v_{\Sf{z}, i}^{-, t}
		+
		|m_{\Sf{z}, i}^{-, t}|^{2}
		a_{\Sf{x}, i}^{-, 2, t}
	}
	+
	\frac{
		(m_{\Sf{z}, i}^{-, t})^{\star}
		\bs{A}_{i} \bs{m}_{\Sf{x}, i}^{+, 1, t}
	}{
		v_{\Sf{z}, i}^{-, t}
		+
		|m_{\Sf{z}, i}^{-, t}|^{2}
		a_{\Sf{x}, i}^{-, 1, t}
	}
]
$\\
$
\bs{\Lambda}_{\tilde{\Sf{x}}}^{-, t}
=
\sum_{i = 1}^{M}{}[
	\frac{
		\bs{A}_{i}^{\Cts}
		\bs{m}_{\Sf{x}, i}^{+, 2, t}
		(\bs{m}_{\Sf{x}, i}^{+, 2, t})^{\Cts}
		\bs{A}_{i}
	}{
		v_{\Sf{z}, i}^{-, t}
		+
		|m_{\Sf{z}, i}^{-, t}|^{2}
		a_{\Sf{x}, i}^{-, 2, t}
	}
	+
	\frac{
		\bs{A}_{i} \bs{m}_{\Sf{x}, i}^{+, 1, t}
		(\bs{m}_{\Sf{x}, i}^{+, 1, t})^{\Cts}
		\bs{A}_{i}^{\Cts}
	}{
		v_{\Sf{z}, i}^{-, t}
		+
		|m_{\Sf{z}, i}^{-, t}|^{2}
		a_{\Sf{x}, i}^{-, 1, t}
	}
]
$\\
$
(
	\hat{\bs{m}}_{\tilde{\Sf{x}}}^{-, t},
	\hat{\bs{C}}_{\tilde{\Sf{x}}}^{-, t}
)
=
\Lex[
	\bs{b}_{\tilde{\Sf{x}}}^{-, t},
	\bs{\Lambda}_{\tilde{\Sf{x}}}^{-, t},
	\bs{m}_{\tilde{\Sf{x}}}^{+, t},
	\bs{v}_{\tilde{\Sf{x}}}^{+, t}
]
$\\
$
\hat{\bs{v}}_{\tilde{\Sf{x}}}^{-, t}
=
\diag(\hat{\bs{C}}_{\tilde{\Sf{x}}}^{-, t})
$\\
$
(
	\bs{m}_{\tilde{\Sf{x}}}^{-, t},
	\bs{v}_{\tilde{\Sf{x}}}^{-, t}
)=
\Ext[
	\hat{\bs{m}}_{\tilde{\Sf{x}}}^{-, t},
	\hat{\bs{v}}_{\tilde{\Sf{x}}}^{-, t},
	\bs{m}_{\tilde{\Sf{x}}}^{+, t},
	\bs{v}_{\tilde{\Sf{x}}}^{+, t}
]
$\\
$
(
	\hat{\bs{m}}_{\tilde{\Sf{x}}}^{+, t + 1},
	\hat{\bs{v}}_{\tilde{\Sf{x}}}^{+, t + 1}
)=
\Mean[
	\tilde{\bs{x}} |
	\bs{m}_{\tilde{\Sf{x}}}^{-, t},
	\bs{v}_{\tilde{\Sf{x}}}^{-, t}
]
$\\
$
(
	\bs{m}_{\tilde{\Sf{x}}}^{+, t + 1},
	\bs{v}_{\tilde{\Sf{x}}}^{+, t + 1}
)
=
\Ext[
	\hat{\bs{m}}_{\tilde{\Sf{x}}}^{+, t + 1},
	\hat{\bs{v}}_{\tilde{\Sf{x}}}^{+, t + 1},
	\bs{m}_{\tilde{\Sf{x}}}^{-, t},
	\bs{v}_{\tilde{\Sf{x}}}^{-, t}
]
$\\
$
(
	\hat{\bs{m}}_{\Sf{x}}^{+, t + 1},
	\hat{\bs{C}}_{\Sf{x}}^{+, t + 1}
)
=
\Lex[
	\bs{b}_{\tilde{\Sf{x}}}^{-, t},
	\bs{\Lambda}_{\tilde{\Sf{x}}}^{-, t},
	\bs{m}_{\tilde{\Sf{x}}}^{+, t + 1},
	\bs{v}_{\tilde{\Sf{x}}}^{+, t + 1}
]
$\\
$
(
	\bs{m}_{\Sf{x}, i}^{+, 1, t + 1},
	\bs{C}_{\Sf{x}, i}^{+, 1, t + 1}
)
=
\Pex[
	\hat{\bs{m}}_{\Sf{x}}^{+, t + 1},
	\hat{\bs{C}}_{\Sf{x}}^{+, t + 1},
	m_{\Sf{z}, i}^{-, t},
	v_{\Sf{z}, i}^{-, t},
	a_{\Sf{x}, i}^{-, 2, t},
	(\bs{m}_{\Sf{x}, i}^{+, 2, t})^{\Cts} \bs{A}_{i}
]
$\\
$
(
	\bs{m}_{\Sf{x}, i}^{+, 2, t + 1},
	\bs{C}_{\Sf{x}, i}^{+, 2, t + 1}
)
=
\Pex[
	\hat{\bs{m}}_{\Sf{x}}^{+, t + 1},
	\hat{\bs{C}}_{\Sf{x}}^{+, t + 1},
	(m_{\Sf{z}, i}^{-, t})^{\star},
	v_{\Sf{z}, i}^{-, t},
	a_{\Sf{x}, i}^{-, 1, t},
	(\bs{m}_{\Sf{x}, i}^{+, 1, t})^{\Cts}
	\bs{A}_{i}^{\Cts}
]
$\\
$
(
	m_{\Sf{z}, i}^{+, t + 1},
	v_{\Sf{z}, i}^{+, t + 1}
)
=
\Ez[
	(\bs{m}_{\Sf{x}, i}^{+, 2, t + 1})^{\Cts}
	\bs{A}_{i},
	\bs{A}_{i}^{\Cts}
	\bs{C}_{\Sf{x}, i}^{+, 2, t + 1}
	\bs{A}_{i},	
	\bs{m}_{\Sf{x}, i}^{+, 1, t + 1},
	\bs{C}_{\Sf{x}, i}^{+, 1, t + 1}
]
$\\
End\\
Output:
$\hat{\bs{m}}_{\tilde{\Sf{x}}}^{+, T + 1}$,
$\hat{\bs{v}}_{\tilde{\Sf{x}}}^{+, T + 1}$
\end{algorithmic}
\end{breakablealgorithm}

After all computations, we have completed the derivation of the QMP algorithm and summarize the iterations of the QMP algorithm, as described in Algo. \ref{Algo:QMP}, which operates in an iterative manner and organizes its message passing in two directions, including one for the forward and the reverse.
We define some notations for ease as below:
\begin{itemize}
\item $
(\hat{\bs{m}}, \hat{\bs{v}})
=
\Mean[\bs{a} | \bs{m}, \bs{v}]
$ denotes the posterior mean and variance of $\bs{a}$, respectively, where the expectations are taken w.r.t. $
\frac{
	\Sf{f}(\bs{a}) \NormalC(\bs{a} | \bs{m}, \bs{v})
}{
	\int{\dd \bs{a}}\,
	\Sf{f}(\bs{a}) \NormalC(\bs{a} | \bs{m}, \bs{v})
}
$;

\item $
(\bs{M}_{2}, \bs{V}_{2})
=
\Ext[
	\hat{\bs{M}},\hat{\bs{V}},
	\bs{M}_{1},\bs{V}_{1}
]
$ represents the extrinsic mean and variance given as $
\bs{V}_{2}
=
\bs{1} \oslash (
	\bs{1} \oslash \hat{\bs{V}}
	-
	\bs{1} \oslash \bs{V}_{1}
)
$ and $
\bs{M}_{2}
=
\bs{V}_{2} \odot (
	\hat{\bs{M}} \oslash \hat{\bs{V}}
	-
	\bs{M}_{1} \oslash \bs{V}_{1}
)
$;

\item $
(\hat{\bs{m}}_{\Sf{x}}, \hat{\bs{C}}_{\Sf{x}})
=
\Lex[
	\bs{b}_{\Sf{x}}, \bs{\Lambda}_{\Sf{x}},
	\bs{m}_{\Sf{x}}, \bs{v}_{\Sf{x}}
]
$ indicates the posterior mean and variance of a LMMSE estimation of $\Sf{x}$ given as $
\hat{\bs{C}}_{\Sf{x}}
=
[
	\bs{\Lambda}_{\Sf{x}}
	+
	\Diag(\bs{1} \oslash \bs{v}_{\Sf{x}})
]^{- 1}
$ and $
\hat{\bs{m}}_{\Sf{x}}
=
\hat{\bs{C}}_{\Sf{x}} (
	\bs{b}_{\Sf{x}}
	+
	\bs{m}_{\Sf{x}} \oslash \bs{v}_{\Sf{x}}
)
$;

\item $
(\bs{m}_{\Sf{x}}, \bs{C}_{\Sf{x}})
=
\Pex[
	\hat{\bs{m}}_{\Sf{x}}, \hat{\bs{C}}_{\Sf{x}},
	m_{\Sf{z}}, v_{\Sf{z}},
	a_{\Sf{h}}, \bs{\Mbar}_{\Sf{h}}
]
$ follows the prior mean and variance of $\Sf{x}$ given as $
\bs{C}_{\Sf{x}}
=
\hat{\bs{C}}_{\Sf{x}}
+
\frac{
	\hat{\bs{C}}_{\Sf{x}} \bs{\Mbar}_{\Sf{h}}^{\Cts}
	\bs{\Mbar}_{\Sf{h}} \hat{\bs{C}}_{\Sf{x}}
}{
	v_{\Sf{z}}
	+
	|m_{\Sf{z}}|^{2}
	a_{\Sf{h}}
	-
	\bs{\Mbar}_{\Sf{h}} \hat{\bs{C}}_{\Sf{x}}
	\bs{\Mbar}_{\Sf{h}}^{\Cts}
}
$ and $
\bs{m}_{\Sf{x}}
=
\hat{\bs{m}}_{\Sf{x}}
-
\frac{
	m_{\Sf{z}}
	-
	\bs{\Mbar}_{\Sf{h}} \hat{\bs{m}}_{\Sf{x}}
}{
	v_{\Sf{z}}
	+
	|m_{\Sf{z}}|^{2}
	a_{\Sf{h}}
	-
	\bs{\Mbar}_{\Sf{h}} \hat{\bs{C}}_{\Sf{x}}
	\bs{\Mbar}_{\Sf{h}}^{\Cts}
}
\hat{\bs{C}}_{\Sf{x}} \bs{\Mbar}_{\Sf{h}}^{\Cts}
$;

\item $
(m_{\Sf{z}}, v_{\Sf{z}})
=
\Ez[
	\bs{\Mbar}_{\Sf{h}}, \bs{C}_{\Sf{h}},
	\bs{m}_{\Sf{x}}, \bs{C}_{\Sf{x}}
]
$ follows the prior mean and variance of $\Sf{z}$ given as $
v_{\Sf{z}}
=
\Trace(
	\bs{C}_{\Sf{h}} \bs{C}_{\Sf{x}}
)
+
\bs{\Mbar}_{\Sf{h}}
\bs{C}_{\Sf{x}}
\bs{\Mbar}_{\Sf{h}}^{\Cts}
+
\bs{m}_{\Sf{x}}^{\Cts}
\bs{C}_{\Sf{h}}
\bs{m}_{\Sf{x}}
$ and $
m_{\Sf{z}}
=
\bs{\Mbar}_{\Sf{h}} \bs{m}_{\Sf{x}}
$.
\end{itemize}

\section{State Evolution of QMP}
In this section, we show the state evolution of the QMP algorithm, which illustrates that the asymptotic MSE performance of the QMP algorithm can be fully characterized via a set of simple one-dimensional equations called state evolution (SE) under the large system limit.
In the literature, Bayati et al. rigorously proved the SE of the original AMP, and later Rangan provided its GAMP extension, more recently He et al. presented the SE for GEC-SR algorithm.
The derivation here is a heuristic proof in the mathematical aspect.
However the SE could precisely capture the per iteration MSE behavior of the QMP algorithm.

\textbf{Step 1)}:
Following the convenient of SE, the randomness of variances, which arises from $\tilde{\bs{x}}$ and $\bs{w}$, should be averaged, and some assumptions are illustrated as that all mean variables
$\hat{\bs{m}}_{\Sf{z}}^{-, t}$,
$\bs{m}_{\Sf{z}}^{-, t}$,
$\hat{\bs{m}}_{\tilde{\Sf{x}}}^{-, t}$,
$\bs{m}_{\tilde{\Sf{x}}}^{-, t}$,
$\hat{\bs{m}}_{\tilde{\Sf{x}}}^{+, t + 1}$,
$\bs{m}_{\tilde{\Sf{x}}}^{+, t + 1}$,
$\hat{\bs{m}}_{\Sf{x}}^{+, t + 1}$,
$\bs{m}_{\Sf{x},i}^{+, 1, t + 1}$,
$\bs{m}_{\Sf{x}, i}^{+, 2, t + 1}$, and
$\bs{m}_{\Sf{z}}^{+, t + 1}$
in Algo. \ref{Algo:QMP} empirically converge to some scalar variables, respectively, e.g.,
\begin{align}
\lim \limits_{
	M, N \rightarrow \infty
}(
	\hat{\bs{m}}_{\Sf{z}}^{-, t},
	\bs{m}_{\Sf{z}}^{-, t},
	\hat{\bs{m}}_{\tilde{\Sf{x}}}^{-, t},
	\bs{m}_{\tilde{\Sf{x}}}^{-, t}
)
\doteq &
(
	\hat{m}_{\Sf{z}}^{-, t},
	m_{\Sf{z}}^{-, t},
	\hat{m}_{\tilde{\Sf{x}}}^{-, t},
	m_{\tilde{\Sf{x}}}^{-, t}
)
,\label{Eq:SE:M_s}\\
\lim \limits_{
	M, N \rightarrow \infty
}(
	\hat{\bs{m}}_{\tilde{\Sf{x}}}^{+, t + 1},
	\bs{m}_{\tilde{\Sf{x}}}^{+, t + 1}
)
\doteq &
(
	\hat{m}_{\tilde{\Sf{x}}}^{+, t + 1},
	m_{\tilde{\Sf{x}}}^{+, t + 1}
)
,\label{Eq:SE:M_p_tx}\\
\lim \limits_{
	M, N \rightarrow \infty
}(
	\hat{\bs{m}}_{\Sf{x}}^{+, t + 1},
	\bs{m}_{\Sf{x}, i}^{+, 1, t + 1},
	\bs{m}_{\Sf{x}, i}^{+, 2, t + 1},
	\bs{m}_{\Sf{z}, i}^{+, t + 1}
)
\doteq &
(
	\hat{m}_{\Sf{x}}^{+, t + 1},
	m_{\Sf{x}}^{+, 1, t + 1},
	m_{\Sf{x}}^{+, 2, t + 1},
	m_{\Sf{z}}^{+, t + 1}
)
,\label{Eq:SE:M_p_xz}
\end{align}
and
$\hat{\bs{v}}_{\Sf{z}}^{-, t}$,
$\bs{v}_{\Sf{z}}^{-, t}$,
$a_{\Sf{x}, i}^{-, 1, t}$,
$a_{\Sf{x}, i}^{-, 2, t}$,
$\hat{\bs{v}}_{\tilde{\Sf{x}}}^{-, t}$,
$\bs{v}_{\tilde{\Sf{x}}}^{-, t}$,
$\hat{\bs{v}}_{\tilde{\Sf{x}}}^{+, t + 1}$,
$\bs{v}_{\tilde{\Sf{x}}}^{+, t + 1}$,
$\hat{\bs{C}}_{\Sf{x}}^{+, t + 1}$,
$\bs{C}_{\Sf{x}, i}^{+, 1, t + 1}$,
$\bs{C}_{\Sf{x}, i}^{+, 2, t + 1}$, and
$\bs{v}_{\Sf{z}}^{+, t + 1}$
in the QMP algorithm also empirically converge to
\begin{align}
\hat{\mathtt{v}}^{-, t}_{\Sf{z}}
= &
\frac{1}{M}
\sum_{i = 1}^{M}{}
\hat{v}^{-, t}_{\Sf{z}, i}
\doteq
\Mean_{y, m^{+, t}_{\Sf{z}}}[
	\hat{v}^{-, t}_{\Sf{z}}
]
,\label{Eq:SE:Hv_s_z}\\
\mathtt{v}^{-, t}_{\Sf{z}}
= &
\frac{1}{M}
\sum_{i = 1}^{M}{}
v^{-, t}_{\Sf{z}, i}
\doteq
\Mean_{y, m^{+, t}_{\Sf{z}}}[
	v^{-,t}_{\Sf{z}}
]
,\label{Eq:SE:v_s_z}\\
\mathtt{a}_{\Sf{x}}^{-, 1, t}
= &
\frac{1}{M}
\sum_{i = 1}^{M}{}
a_{\Sf{x}, i}^{-, 1, t}
\doteq
\Mean_{m_{\Sf{x}}^{+, 1, t}}[
	a_{\Sf{x}}^{-, 1, t}
]
,\label{Eq:SE:a_s1_x}\\
\mathtt{a}_{\Sf{x}}^{-, 2, t}
= &
\frac{1}{M}
\sum_{i = 1}^{M}{}
a_{\Sf{x}, i}^{-, 2, t}
\doteq
\Mean_{m_{\Sf{x}}^{+, 2, t}}[
	a_{\Sf{x}}^{-, 2, t}
]
,\label{Eq:SE:a_s2_x}\\
\hat{\mathtt{v}}_{\tilde{\Sf{x}}}^{-, t}
= &
\frac{1}{N}
\sum_{i = 1}^{N}{}
\hat{v}_{\tilde{\Sf{x}}, i}^{-, t}
\doteq
\Mean_{
	m_{\Sf{x}}^{+, 1, t},
	m_{\Sf{x}}^{+, 2, t}
}[
	\hat{v}_{\tilde{\Sf{x}}}^{-, t}
]
,\label{Eq:SE:Hv_s_tx}\\
\mathtt{v}_{\tilde{\Sf{x}}}^{-, t}
= &
\frac{1}{N}
\sum_{i = 1}^{N}{}
v_{\tilde{\Sf{x}}, i}^{-, t}
\doteq
\Mean_{
	m_{\Sf{x}}^{+, 1, t},
	m_{\Sf{x}}^{+, 2, t}
}[
	v_{\tilde{\Sf{x}}}^{-, t}
]
,\label{Eq:SE:v_s_tx}\\
\hat{\mathtt{v}}_{\tilde{\Sf{x}}}^{+, t + 1}
= &
\frac{1}{N}
\sum_{i = 1}^{N}{}
\hat{v}_{\tilde{\Sf{x}}, i}^{+, t + 1}
\doteq
\Mean_{m_{\tilde{\Sf{x}}}^{-, t}}[
	\hat{v}_{\tilde{\Sf{x}}}^{+, t + 1}
]
,\label{Eq:SE:Hv_p_tx}\\
\mathtt{v}^{+, t + 1}_{\tilde{\Sf{x}}}
= &
\frac{1}{N}
\sum_{i = 1}^{N}{}
v_{\tilde{\Sf{x}}, i}^{+, t + 1}
\doteq
\Mean_{m_{\tilde{\Sf{x}}}^{-, t}}[
	v^{+, t + 1}_{\tilde{\Sf{x}}}
]
,\label{Eq:SE:v_p_tx}\\
\hat{\mathtt{c}}_{\Sf{x}}^{+, t + 1}
= &
\frac{1}{N}
\Trace(\hat{\bs{C}}_{\Sf{x}}^{+, t + 1})
\doteq
\Mean_{
	m_{\Sf{x}}^{+, 1, t},
	m_{\Sf{x}}^{+, 2, t}
}[
	\hat{c}_{\Sf{x}}^{+, t + 1}
]
,\label{Eq:SE:Hc_p_x}\\
\mathtt{c}_{\Sf{x}}^{+, 1, t + 1}
= &
\frac{1}{M N}
\sum_{i = 1}^{N}{}
\Trace(\bs{C}_{\Sf{x}, i}^{+, 1, t + 1})
\doteq
\Mean_{
	m_{\Sf{x}}^{+, 1, t},
	m_{\Sf{x}}^{+, 2, t}
}[
	c_{\Sf{x}}^{+, 1, t + 1}
]
,\label{Eq:SE:c_p1_x}\\
\mathtt{c}_{\Sf{x}}^{+, 2, t + 1}
= &
\frac{1}{M N}
\sum_{i = 1}^{N}{}
\Trace(\bs{C}_{\Sf{x}, i}^{+, 2, t + 1})
\doteq
\Mean_{
	m_{\Sf{x}}^{+, 1, t},
	m_{\Sf{x}}^{+, 2, t}
}[
	c_{\Sf{x}}^{+, 2, t + 1}
]
,\label{Eq:SE:c_p2_x}\\
\mathtt{v}_{\Sf{z}}^{+, t + 1}
= &
\frac{1}{M}
\sum_{i = 1}^{M}{}
v_{\Sf{z}, i}^{+, t + 1}
\doteq
\Mean_{
	m_{\Sf{x}}^{+, 1, t + 1},
	m_{\Sf{x}}^{+, 2, t + 1}
}[
	v_{\Sf{z}}^{+, t + 1}
]
,\label{Eq:SE:v_p_z}
\end{align}
where we use typewriter-style small letters for variance variables in SE (i.e, $\mathtt{v}$, $\mathtt{c}$ and $\mathtt{a}$).

\TB{Step 2)}: For Line 13,
$m^{-, t}_{\tilde{\Sf{x}}}$
can be systematically decoupled by a scalar channel as
\begin{align}
m^{-, t}_{\tilde{\Sf{x}}}
= &
\tilde{x} + n
,\label{Eq:SE:Decomposition_tx}
\end{align}
where $\tilde{x}$ and $n$ follow
$p(\tilde{x})$
and
$\NormalC(n | 0, \mathtt{v}^{-, t}_{\tilde{\Sf{x}}})$,
respectively.
So we can calculate
$\hat{\mathtt{v}}^{+, t + 1}_{\tilde{\Sf{x}}}$
and
$\mathtt{v}^{+, t + 1}_{\tilde{\Sf{x}}}$
as 
\begin{align}
\hat{\mathtt{v}}^{+, t + 1}_{\tilde{\Sf{x}}}
= &
\Mean_{m_{\tilde{\Sf{x}}}^{-, t}}[
	\hat{v}_{\tilde{\Sf{x}}}^{+, t + 1}
]
=
T_{\tilde{\Sf{x}}}
-
q^{+, t + 1}_{\tilde{\Sf{x}}}
,\label{Eq:SE:Update:Hv_p_tx}\\
\mathtt{v}^{+, t + 1}_{\tilde{\Sf{x}}}
= &
(
	\frac{1}{
		\hat{\mathtt{v}}^{+, t + 1}_{\tilde{\Sf{x}}}
	}
	-
	\mathtt{v}^{-, t}_{\tilde{\Sf{x}}}
)^{- 1}
,\label{Eq:SE:Update:v_p_tx}
\end{align}
where
$
T_{\tilde{\Sf{x}}}
=
\int{\dd \tilde{x}}\,
\tilde{x}^{2} p(\tilde{x})
$
and
$
q^{+, t + 1}_{\tilde{\Sf{x}}}
=
\int{\dd \zeta}\,
\frac{
	[
		\int{\dd \tilde{x}}\,
		\tilde{x} p(\tilde{x})
		\NormalC(
			\tilde{x} | \zeta,
			\mathtt{v}^{-, t}_{\tilde{\Sf{x}}}
		)
	]^{2}
}{
	\int{\dd \tilde{x}}\,
	p(\tilde{x})
	\NormalC(
		\tilde{x} | \zeta,
		\mathtt{v}^{-, t}_{\tilde{\Sf{x}}}
	)
}
$. 

\TB{Step 3)}: According to \eqref{Eq:Update_C_p1t1_xi}, we continue to handle Line 16 as
\begin{align}
\bs{C}_{\Sf{x}, i}^{+, 1, t + 1}
= &
[
	\bs{\Lambda}_{\Sf{x} \Bs i}^{-, 1, t}
	+
	\Diag(\bs{1} \oslash
	\bs{v}_{\tilde{\Sf{x}}}^{+, t + 1})
]^{- 1}
\nonumber\\
= &
[
	\sum_{k \neq i}^{M}{}
	\frac{
		\bs{A}_{k}^{\Cts}
		\bs{m}_{\Sf{x}, k}^{+, 2, t}
		(\bs{m}_{\Sf{x}, k}^{+, 2, t})^{\Cts}
		\bs{A}_{k}
	}{
		v_{\Sf{z}, k}^{-, t}
		+
		|m_{\Sf{z}, k}^{-, t}|^{2}
		a_{\Sf{x}, k}^{-, 2, t}
	}
	+
	\sum_{k = 1}^{M}{}
	\frac{
		\bs{A}_{k} \bs{m}_{\Sf{x}, k}^{+, 1, t}
		(\bs{m}_{\Sf{x}, k}^{+, 1, t})^{\Cts}
		\bs{A}_{k}^{\Cts}
	}{
		v_{\Sf{z}, k}^{-, t}
		+
		|m_{\Sf{z}, k}^{-, t}|^{2}
		a_{\Sf{x}, k}^{-, 1, t}
	}
	+
	\Diag(\bs{1} \oslash
	\bs{v}_{\tilde{\Sf{x}}}^{+, t + 1})
]^{- 1}
,\label{Eq:SE:Update:C_p1_xi}
\end{align}
and
$c^{+, t + 1}_{\Sf{x}}$
denotes as
\begin{align}
c_{\Sf{x}}^{+, 1, t + 1}
= &
\frac{1}{N}
\Trace(
	\bs{C}_{\Sf{x}}^{+, 1, t + 1}
)
\nonumber\\
= &
\frac{1}{N}
\Trace[
	\frac{1}{
		\mathtt{w}^{-, 2, t}_{\Sf{x}}
	}
	\sum_{k \neq 1}^{M}{}
	\bs{A}_{k}^{\Cts}
	\bs{m}_{\Sf{x}, k}^{+, 2, t}
	(\bs{m}_{\Sf{x}, k}^{+, 2, t})^{\Cts}
	\bs{A}_{k}
	+
	\frac{1}{
		\mathtt{w}^{-, 1, t}_{\Sf{x}}
	}
	\sum_{k = 1}^{M}{}
	\bs{A}_{k} \bs{m}_{\Sf{x}, k}^{+, 1, t}
	(\bs{m}_{\Sf{x}, k}^{+, 1, t})^{\Cts}
	\bs{A}_{k}^{\Cts}
	+
	\frac{1}{
		\mathtt{v}^{+, t + 1}_{\tilde{\Sf{x}}}
	} \Eye
]^{-1}
,\label{Eq:SE:Update:c_p1_x}
\end{align}
where
$
\mathtt{w}^{-, t}_{\Sf{z}}
=
T_{\Sf{z}} + \mathtt{v}^{-, t}_{\Sf{z}}
$,
$
\mathtt{w}^{-, 1, t}_{\Sf{x}}
=
\mathtt{v}^{-, t}_{\Sf{z}}
+
\mathtt{w}^{-, t}_{\Sf{z}}
\mathtt{a}^{-, 1, t}_{\Sf{x}}
$,
and
$
\mathtt{w}^{-, 2, t}_{\Sf{x}}
=
\mathtt{v}^{-, t}_{\Sf{z}}
+
\mathtt{w}^{-, t}_{\Sf{z}}
\mathtt{a}^{-, 2, t}_{\Sf{x}}
$.
\eqref{Eq:SE:Update:c_p1_x} is transformed into some equations as follows
\begin{align}
c_{\Sf{x}}^{+, 1, t + 1}
= &
\Mean_{\lambda}[
	\frac{1}{
		\lambda
		+
		\frac{1}{
			\mathtt{v}^{+, t + 1}_{\tilde{\Sf{x}}}
		}
	}
]
,\label{Eq:SE:Update:c_p1_x_reformula}
\end{align}
where
$
\lambda
=
\Eig[
	\frac{1}{
		\mathtt{w}^{-, 2, t}_{\Sf{x}}
	}
	\sum_{k \neq 1}^{M}{}
	\bs{A}_{k}^{\Cts}
	\bs{m}_{\Sf{x}, k}^{+, 2, t}
	(\bs{m}_{\Sf{x}, k}^{+, 2, t})^{\Cts}
	\bs{A}_{k}
	+
	\frac{1}{
		\mathtt{w}^{-, 1, t}_{\Sf{x}}
	}
	\sum_{k = 1}^{M}{}
	\bs{A}_{k} \bs{m}_{\Sf{x}, k}^{+, 1, t}
	(\bs{m}_{\Sf{x}, k}^{+, 1, t})^{\Cts}
	\bs{A}_{k}^{\Cts}
]
$.
Finally
$\mathtt{c}_{\Sf{x}}^{+, 1, t + 1}$
denotes as
\begin{align}
\mathtt{c}_{\Sf{x}}^{+, 1, t + 1}
 = &
\Mean_{
	m_{\Sf{x}}^{+, 1, t},
	m_{\Sf{x}}^{+, 2, t}
}[
	c_{\Sf{x}}^{+, 1, t + 1}
]
=
\Mean_{
	m_{\Sf{x}}^{+, 1, t},
	m_{\Sf{x}}^{+, 2, t}
}(
	\Mean_{\lambda}[
		\frac{1}{
			\lambda
			+
			\frac{1}{
				\mathtt{v}^{+, t + 1}_{\tilde{\Sf{x}}}
			}
		}
	]	
)
.\label{Eq:SE:Update:c_p1_x_SE}
\end{align}
We need to analyze the prior distributions
$p(m_{\Sf{x}}^{+, 1, t})$
and
$p(m_{\Sf{x}}^{+, 2, t})$,
which is the randomness of
$\mathtt{c}_{\Sf{x}}^{+, 1, t + 1}$.
However, the exact prior distributions of
$m_{\Sf{x}}^{+, 1, t}$
and
$m_{\Sf{x}}^{+, 2, t}$
is mostly prohibitive, so we try to analyze the corresponding message via Markov chain, and obtain the approximate prior distributions of
$m_{\Sf{x}}^{+, 1, t}$
and
$m_{\Sf{x}}^{+, 2, t}$.

For Line 16, we interpret message
$
\NormalC(
	x | m^{+, t + 1}_{\Sf{x}},
	\mathtt{c}^{+, t + 1}_{\Sf{x}}
)
$
and obtain the approximate prior distribution of
$m^{+, 1, t + 1}_{\Sf{x}}$
as
\begin{align}
p(x)
= &
\NormalC(
	x | m_{\tilde{\Sf{x}}}^{+, t + 1},
	\mathtt{v}_{\tilde{\Sf{x}}}^{+, t + 1}
)
=
\int{\dd m^{+, 1, t + 1}_{\Sf{x}}}\,
p(m^{+, 1, t + 1}_{\Sf{x}})
\NormalC(
	x | m^{+, 1, t + 1}_{\Sf{x}},
	\mathtt{c}^{+, 1, t + 1}_{\Sf{x}}
)
,\label{Eq:SE:Update:P_x}\\
p(m^{+, 1, t + 1}_{\Sf{x}})
= &
\NormalC[
	m^{+, 1, t + 1}_{\Sf{x}} |
	m_{\tilde{\Sf{x}}}^{+, t + 1},
	\mathtt{v}_{\tilde{\Sf{x}}}^{+, t + 1}
	-
	\mathtt{c}^{+, 1, t + 1}_{\Sf{x}}
]
,\label{Eq:SE:Update:P_m_p1_x}
\end{align}
where
$m_{\tilde{\Sf{x}}}^{+, t + 1}$
is hardly to be modeled, so we only use the Monte Carlo method \cite{rubinstein2016simulation,metropolis1949monte} to generate
$m_{\tilde{\Sf{x}}}^{+, t + 1}$.

\TB{Step 4)}: Line 17 are similarly analyzed as
\begin{align}
c_{\Sf{x}}^{+, 2, t + 1}
= &
\Mean_{\lambda}[
	\frac{1}{
		\lambda
		+
		\frac{1}{
			\mathtt{v}^{+, t + 1}_{\tilde{\Sf{x}}}
		}
	}
]
,\label{Eq:SE:Update:c_p2_x_reformula}\\
\mathtt{c}_{\Sf{x}}^{+, 2, t + 1}
 = &
\Mean_{
	m_{\Sf{x}}^{+, 1, t},
	m_{\Sf{x}}^{+, 2, t}
}[
	c_{\Sf{x}}^{+, 2, t + 1}
]
=
\Mean_{
	m_{\Sf{x}}^{+, 1, t},
	m_{\Sf{x}}^{+, 2, t}
}(
	\Mean_{\lambda}[
		\frac{1}{
			\lambda
			+
			\frac{1}{
				\mathtt{v}^{+, t + 1}_{\tilde{\Sf{x}}}
			}
		}
	]	
)
,\label{Eq:SE:Update:c_p2_x_SE}\\
p(m^{+, 2, t + 1}_{\Sf{x}})
= &
\NormalC[
	m^{+, 2, t + 1}_{\Sf{x}} |
	m_{\tilde{\Sf{x}}}^{+, t + 1},
	\mathtt{v}_{\tilde{\Sf{x}}}^{+, t + 1}
	-
	\mathtt{c}^{+, 2, t + 1}_{\Sf{x}}
]
,\label{Eq:SE:Update:P_m_p2_x}
\end{align}
where
$
\lambda
=
\Eig[
	\frac{1}{
		\mathtt{w}^{-, 2, t}_{\Sf{x}}
	}
	\sum_{k = 1}^{M}{}
	\bs{A}_{k}^{\Cts}
	\bs{m}_{\Sf{x}, k}^{+, 2, t}
	(\bs{m}_{\Sf{x}, k}^{+, 2, t})^{\Cts}
	\bs{A}_{k}
	+
	\frac{1}{
		\mathtt{w}^{-, 1, t}_{\Sf{x}}
	}
	\sum_{k \neq 1}^{M}{}
	\bs{A}_{k} \bs{m}_{\Sf{x}, k}^{+, 1, t}
	(\bs{m}_{\Sf{x}, k}^{+, 1, t})^{\Cts}
	\bs{A}_{k}^{\Cts}
]
$
.

\TB{Step 5)}: Now we denote Line 18 as
\begin{align}
v_{\Sf{z}}^{+, t + 1}
= &
\frac{1}{M}
\sum_{i = 1}^{M}{}
v_{\Sf{z}, i}^{+, t + 1}
\nonumber\\
= &
\mathtt{c}^{+, 1, t + 1}_{\Sf{x}}
\mathtt{c}^{+, 2, t + 1}_{\Sf{x}}
\Trace(
	\frac{1}{M}
	\sum_{i = 1}^{M}{}
	\bs{A}_{i} \bs{A}_{i}^{\Cts}
)
+
\mathtt{c}^{+, 1, t + 1}_{\Sf{x}}
\Trace[
	\frac{1}{M}
	\sum_{i = 1}^{M}{}
	\bs{A}_{i}^{\Cts}
	\bs{m}_{\Sf{x}}^{+, 2, t + 1}
	(\bs{m}_{\Sf{x}}^{+, 2, t + 1})^{\Cts}
	\bs{A}_{i}
]
+
\nonumber\\
&
\mathtt{c}^{+, 2, t + 1}_{\Sf{x}}
\Trace[
	\frac{1}{M}
	\sum_{i = 1}^{M}{}
	\bs{A}_{i}
	\bs{m}_{\Sf{x}}^{+, 1, t + 1}
	(\bs{m}_{\Sf{x}}^{+, 1, t + 1})^{\Cts}
	\bs{A}_{i}^{\Cts}
]
,\label{Eq:SE:Update:v_p_z}\\
\mathtt{v}_{\Sf{z}}^{+, t + 1}
= &
\Mean_{
	m_{\Sf{x}}^{+, 1, t + 1},
	m_{\Sf{x}}^{+, 2, t + 1}
}[
	v_{\Sf{z}}^{+, t + 1}
]
.\label{Eq:SE:Update:v_p_z_SE}
\end{align}

\TB{Step 6)}: Using central limit theorem, we firstly approximate the prior distribution of $z$ as
$\NormalC(z | m_{\Sf{z}}, T_{\Sf{z}})$
with
\begin{align}
m_{\Sf{z}}
= &
\int{\dd z}\,
z p(z)
=
0
,\label{Eq:SE:Update:m_z}\\
T_{\Sf{z}}
= &
\int{\dd z}\,
z^{2} p(z)
=
N^{2} T_{\Sf{a}} T_{\tilde{\Sf{x}}}
.\label{Eq:SE:Update:T_z}
\end{align}

\begin{figure}[!t]
\centering
\includegraphics[width=0.7\textwidth]{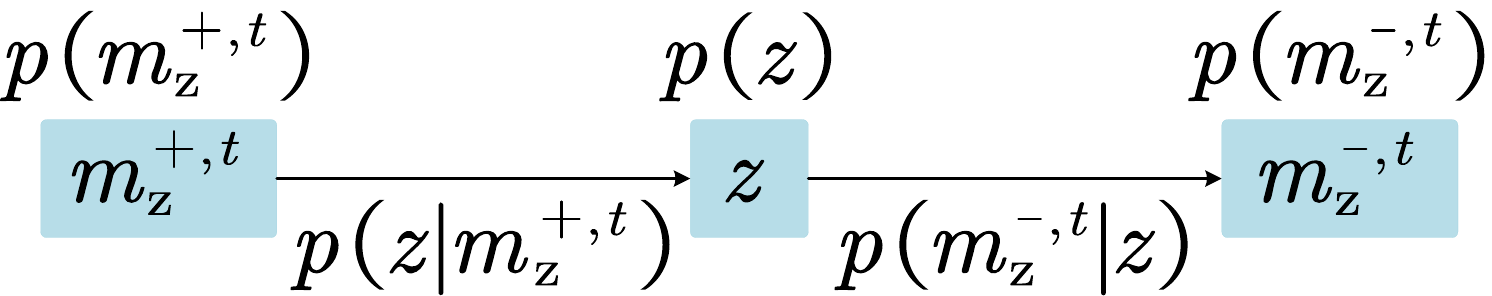}
\caption{
	The Markov chain sequence: $
	m^{+, t}_{\Sf{z}}
	\rightarrow
	z
	\rightarrow
	m^{-, t}_{\Sf{z}}
	$.
}
\label{Fig:Markov_Chain_Z}
\end{figure}

For Lines 4-5, we similarly interpret the messages
$
\NormalC(
	z | m^{+, t}_{\Sf{z}},
	\mathtt{v}^{+, t}_{\Sf{z}}
)
$
and
$
\NormalC(
	z | m^{-, t}_{\Sf{z}},
	\mathtt{v}^{-, t}_{\Sf{z}}
)
$
as a two-steps Markov chain, as depicted in Fig. \ref{Fig:Markov_Chain_Z}, and obtain the approximate prior distributions
$p(m^{+, t}_{\Sf{z}})$
and
$p(m^{-, t}_{\Sf{z}})$
as
\begin{align}
p(z)
= &
\NormalC(z | 0, T_{\Sf{z}})
=
\int{\dd m^{+,t}_{\Sf{z}}}\,
p(m^{+, t}_{\Sf{z}})
\NormalC(
	z | m^{+, t}_{\Sf{z}},
	\mathtt{v}^{+, t}_{\Sf{z}}
)
,\label{Eq:SE:Update:P_z}\\
p(m^{+, t}_{\Sf{z}})
= &
\NormalC[
	m^{+, t}_{\Sf{z}} | 0,
	T_{\Sf{z}} - \mathtt{v}^{+, t}_{\Sf{z}}
]
,\label{Eq:SE:Update:P_m_p_z}\\
p(m^{-, t}_{\Sf{z}})
= &
\int{\dd z}\,
p(z)
\NormalC(
	z | m^{-, t}_{\Sf{z}},
	\mathtt{v}^{-, t}_{\Sf{z}}
)
=
\NormalC(
	m^{-, t}_{\Sf{z}} | 0,
	\mathtt{w}^{-, t}_{\Sf{z}}
)
,\label{Eq:SE:Update:P_m_s_z}
\end{align}
where
$
\mathtt{w}^{-, t}_{\Sf{z}}
=
T_{\Sf{z}} + \mathtt{v}^{-, t}_{\Sf{z}}
$.

Combining \eqref{Eq:SE:Update:P_m_s_z} into Lines 4-5, obtains
\begin{align}
\mathtt{\hat{v}}^{-, t}_{\Sf{z}}
= &
\Mean_{
	y, m^{+, t}_{\Sf{z}}
}[
	\hat{v}^{-, t}_{\Sf{z}}
]
=
T_{\Sf{z}} - q^{-, t}_{\Sf{z}}
,\label{Eq:SE:Update:Hv_s_z}\\
\mathtt{v}^{-, t}_{\Sf{z}}
= &
(
	\frac{1}{
		\mathtt{\hat{v}}^{-, t}_{\Sf{z}}
	}
	-
	\frac{1}{
		\mathtt{v}^{+, t}_{\Sf{z}}
	}
)^{-1}
,\label{Eq:SE:Update:v_s_z}
\end{align}
where 
\begin{align}
q^{-, t}_{\Sf{z}}
= &
\int{\dd y \dd \xi}\,
\NormalC(\xi | 0, 1)
\frac{
	|
		\int{\dd z}\,
		z p(y | z)
		\NormalC(
			z |
			\sqrt{
				T_{\Sf{z}}
				-
				\mathtt{v}^{+, t}_{\Sf{z}}
			} \xi,
			\mathtt{v}^{+, t}_{\Sf{z}}
		)
	|^{2}
}{
	\int{\dd z}\,
	p(y | z)
	\NormalC(
		z |
		\sqrt{
			T_{\Sf{z}}
			-
			\mathtt{v}^{+, t}_{\Sf{z}}
		} \xi,
		\mathtt{v}^{+, t}_{\Sf{z}}
	)
}
.\label{Eq:SE:Update:q_s_z}
\end{align}

\TB{Step 7)}: Lines 6-7 are calculated as
\begin{align}
a_{\Sf{x}, i}^{-, 1, t}
= &
\mathtt{c}_{\Sf{x}}^{+, 1, t}
\frac{
	(\bs{m}_{\Sf{x}, i}^{+, 1, t})^{\Cts}
	\bs{A}_{i}^{\Cts}
	\bs{A}_{i}
	\bs{A}_{i}^{\Cts}
	\bs{A}_{i} \bs{m}_{\Sf{x}, i}^{+, 1, t}
}{
	|
		(\bs{m}_{\Sf{x}, i}^{+, 1, t})^{\Cts}
		\bs{A}_{i}^{\Cts}
		\bs{A}_{i} \bs{m}_{\Sf{x}, i}^{+, 1, t}
	|^{2}
}
=
\frac{
	\mathtt{c}_{\Sf{x}}^{+, 1, t}
}{
	(\bs{m}_{\Sf{x}, i}^{+, 1, t})^{\Cts}
	\bs{m}_{\Sf{x}, i}^{+, 1, t}
}
,\label{Eq:SE:Update:a_s1_xi}\\
\mathtt{a}_{\Sf{x}}^{-, 1, t}
= &
\frac{
	\mathtt{c}_{\Sf{x}}^{+, 1, t}
}{
	\frac{1}{M}
	\Trace[
		\bs{M}_{\Sf{x}}^{+, 1, t}
		(\bs{M}_{\Sf{x}}^{+, 1, t})^{\Cts}
	]
}
,\label{Eq:SE:Update:a_s1_x}\\
\mathtt{a}_{\Sf{x}}^{-, 2, t}
= &
\frac{
	\mathtt{c}_{\Sf{x}}^{+, 2, t}
}{
	\frac{1}{M}
	\Trace[
		\bs{M}_{\Sf{x}}^{+, 2, t}
		(\bs{M}_{\Sf{x}}^{+, 2, t})^{\Cts}
	]
}
.\label{Eq:SE:Update:a_s2_x}
\end{align}

\TB{Step 8)}: Lines 8-12 are similarly computed as
\begin{align}
\hat{\mathtt{v}}^{-, t}_{\tilde{\Sf{x}}}
= &
\Mean_{
	m_{\Sf{x}}^{+, 1, t},
	m_{\Sf{x}}^{+, 2, t}
}(
	\Mean_{\lambda}[
		\frac{1}{
			\lambda
			+
			\frac{1}{
				\mathtt{v}^{+, t}_{\tilde{\Sf{x}}}
			}
		}
	]	
)
,\label{Eq:SE:Update:Hv_s_tx}\\
\mathtt{v}^{-, t}_{\tilde{\Sf{x}}}
= &
(
	\frac{1}{
		\mathtt{\hat{v}}^{-, t}_{\tilde{\Sf{x}}}
	}
	-
	\frac{1}{
		\mathtt{v}^{-, t}_{\tilde{\Sf{x}}}
	}
)^{- 1}
,\label{Eq:SE:Update:v_s_tx}
\end{align}
where
$
\lambda
=
\Eig[
	\frac{1}{
		\mathtt{w}^{-, 2, t}_{\Sf{x}}
	}
	\sum_{k = 1}^{M}{}
	\bs{A}_{k}^{\Cts}
	\bs{m}_{\Sf{x}, k}^{+, 2, t}
	(\bs{m}_{\Sf{x}, k}^{+, 2, t})^{\Cts}
	\bs{A}_{k}
	+
	\frac{1}{
		\mathtt{w}^{-, 1, t}_{\Sf{x}}
	}
	\sum_{k = 1}^{M}{}
	\bs{A}_{k} \bs{m}_{\Sf{x}, k}^{+, 1, t}
	(\bs{m}_{\Sf{x}, k}^{+, 1, t})^{\Cts}
	\bs{A}_{k}^{\Cts}
]
$.

So far, we have obtained the explicit expressions of
$\hat{\mathtt{v}}^{-, t}_{\Sf{z}}$,
$\mathtt{v}^{-, t}_{\Sf{z}}$,
$\mathtt{a}_{\Sf{x}}^{-, 1, t}$,
$\mathtt{a}_{\Sf{x}}^{-, 2, t}$,
$\hat{\mathtt{v}}_{\tilde{\Sf{x}}}^{-, t}$,
$\mathtt{v}_{\tilde{\Sf{x}}}^{-, t}$,
$\hat{\mathtt{v}}_{\tilde{\Sf{x}}}^{+, t + 1}$,
$\mathtt{v}^{+, t + 1}_{\tilde{\Sf{x}}}$,
$\hat{\mathtt{c}}_{\Sf{x}}^{+, t + 1}$,
$\mathtt{c}_{\Sf{x}}^{+, 1, t + 1}$,
$\mathtt{c}_{\Sf{x}}^{+, 2, t + 1}$,
$\mathtt{v}_{\Sf{z}}^{+, t + 1}$
and the approximate prior distributions of
$m^{+, 1, t}_{\Sf{x}}$,
$m^{+, 2, t}_{\Sf{x}}$,
and finally summarize the recursive iteration equations in Algo. \ref{Algo:SE:QS_MP}.

\begin{breakablealgorithm}
\setstretch{2}
\caption{The SE of the QMP algorithm}
\label{Algo:SE:QS_MP}
\begin{algorithmic}[1]
\State
Input:
$\bs{A}_{k}$,
$T_{\Sf{a}}$,
$
T_{\tilde{\Sf{x}}}
=
\int{\dd \tilde{x}}\, \tilde{x}^{2} p(\tilde{x})
$,
$T_{\Sf{z}} = N^{2} T_{\Sf{a}} T_{\tilde{\Sf{x}}}$ \\
Initialize:
$\mathtt{v}^{+, 1}_{\Sf{z}}$,
$\mathtt{v}^{+, 1}_{\tilde{\Sf{x}}}$,
$\bs{M}_{\Sf{x}}^{+, 1, 1}$,
$\bs{M}_{\Sf{x}}^{+, 2, 1}$,
$\mathtt{c}_{\Sf{x}}^{+, 1, 1}$,
$\mathtt{c}_{\Sf{x}}^{+, 2, 1}$ \\
Iterate: $t = 1, \cdots, T$ \\
$
q^{-, t}_{\Sf{z}}
=
\int{\dd y \dd \xi}\,
\NormalC(\xi | 0, 1)
\frac{
	|
		\int{\dd z}\,
		z p(y | z)
		\NormalC(
			z |
			\sqrt{
				T_{\Sf{z}}
				-
				\mathtt{v}^{+, t}_{\Sf{z}}
			} \xi,
			\mathtt{v}^{+, t}_{\Sf{z}}
		)
	|^{2}
}{
	\int{\dd z}\,
	p(y | z)
	\NormalC(
		z |
		\sqrt{
			T_{\Sf{z}}
			-
			\mathtt{v}^{+, t}_{\Sf{z}}
		} \xi,
		\mathtt{v}^{+, t}_{\Sf{z}}
	)
}
$ \\
$
\mathtt{\hat{v}}^{-, t}_{\Sf{z}}
=
T_{\Sf{z}} - q^{-, t}_{\Sf{z}}
$ \\
$
\mathtt{v}^{-, t}_{\Sf{z}}
=
(
	\frac{1}{
		\mathtt{\hat{v}}^{-, t}_{\Sf{z}}
	}
	-
	\frac{1}{
		\mathtt{v}^{+, t}_{\Sf{z}}
	}
)^{-1}
$ \\
$
\mathtt{w}^{-, t}_{\Sf{z}}
=
T_{\Sf{z}} + \mathtt{v}^{-, t}_{\Sf{z}}
$ \\
$
\mathtt{a}_{\Sf{x}}^{-, 1, t}
=
\frac{
	\mathtt{c}_{\Sf{x}}^{+, 1, t}
}{
	\frac{1}{M}
	\Trace\{
		\Mean_{
			m_{\Sf{x}}^{+, 1, t}
		}[
			\bs{M}_{\Sf{x}}^{+, 1, t}
			(\bs{M}_{\Sf{x}}^{+, 1, t})^{\Cts}
		]
	\}
}
$ \\
$
\mathtt{a}_{\Sf{x}}^{-, 2, t}
=
\frac{
	\mathtt{c}_{\Sf{x}}^{+, 2, t}
}{
	\frac{1}{M}
	\Trace\{
		\Mean_{
			m_{\Sf{x}}^{+, 2, t}
		}[
			\bs{M}_{\Sf{x}}^{+, 2, t}
			(\bs{M}_{\Sf{x}}^{+, 2, t})^{\Cts}
		]
	\}
}
$ \\
$
\mathtt{w}^{-, 1, t}_{\Sf{x}}
=
\mathtt{v}^{-, t}_{\Sf{z}}
+
\mathtt{w}^{-, t}_{\Sf{z}}
\mathtt{a}^{-, 1, t}_{\Sf{x}}
$ \\
$
\mathtt{w}^{-, 2, t}_{\Sf{x}}
=
\mathtt{v}^{-, t}_{\Sf{z}}
+
\mathtt{w}^{-, t}_{\Sf{z}}
\mathtt{a}^{-, 2, t}_{\Sf{x}}
$ \\
$
\bs{m}_{\Sf{a}, k}^{+, 1, t}
=
\bs{A}_{k}
\bs{m}_{\Sf{x}, k}^{+, 1, t}
$ \\
$
\bs{m}_{\Sf{a}, k}^{+, 2, t}
=
\bs{A}_{k}^{\Cts}
\bs{m}_{\Sf{x}, k}^{+, 2, t}
$ \\
$
\hat{\mathtt{v}}^{-, t}_{\tilde{\Sf{x}}}
=
\Mean_{
	m_{\Sf{x}}^{+, 1, t},
	m_{\Sf{x}}^{+, 2, t}
}(
	\Mean_{\lambda}[
		\frac{1}{
			\lambda
			+
			\frac{1}{
				\mathtt{v}^{+, t}_{\tilde{\Sf{x}}}
			}
		}
	]	
)
$, $
\lambda
=
\Eig[
	\frac{1}{
		\mathtt{w}^{-, 2, t}_{\Sf{x}}
	}
	\bs{M}_{\Sf{a}}^{+, 2, t}
	(\bs{M}_{\Sf{a}}^{+, 2, t})^{\Cts}
	+
	\frac{1}{
		\mathtt{w}^{-, 1, t}_{\Sf{x}}
	}
	\bs{M}_{\Sf{a}}^{+, 1, t}
	(\bs{M}_{\Sf{a}}^{+, 1, t})^{\Cts}
]
$ \\
$
\mathtt{v}^{-, t}_{\tilde{\Sf{x}}}
=
(
	\frac{1}{
		\mathtt{\hat{v}}^{-, t}_{\tilde{\Sf{x}}}
	}
	-
	\frac{1}{
		\mathtt{v}^{-, t}_{\tilde{\Sf{x}}}
	}
)^{- 1}
$ \\
$
q^{+, t + 1}_{\tilde{\Sf{x}}}
=
\int{\dd \zeta}\,
\frac{
	[
		\int{\dd \tilde{x}}\,
		\tilde{x} p(\tilde{x})
		\NormalC(
			\tilde{x} | \zeta,
			\mathtt{v}^{-, t}_{\tilde{\Sf{x}}}
		)
	]^{2}
}{
	\int{\dd \tilde{x}}\,
	p(\tilde{x})
	\NormalC(
		\tilde{x} | \zeta,
		\mathtt{v}^{-, t}_{\tilde{\Sf{x}}}
	)
}
$ \\
$
\hat{\mathtt{v}}^{+, t + 1}_{\tilde{\Sf{x}}}
=
T_{\tilde{\Sf{x}}}
-
q^{+, t + 1}_{\tilde{\Sf{x}}}
$ \\
$
\mathtt{v}^{+, t + 1}_{\tilde{\Sf{x}}}
=
(
	\frac{1}{
		\hat{\mathtt{v}}^{+, t + 1}_{\tilde{\Sf{x}}}
	}
	-
	\mathtt{v}^{-, t}_{\tilde{\Sf{x}}}
)^{- 1}
$ \\
$
\mathtt{c}_{\Sf{x}}^{+, 1, t + 1}
=
\Mean_{
	m_{\Sf{x}}^{+, 1, t},
	m_{\Sf{x}}^{+, 2, t}
}(
	\Mean_{\lambda}[
		\frac{1}{
			\lambda
			+
			\frac{1}{
				\mathtt{v}^{+, t + 1}_{\tilde{\Sf{x}}}
			}
		}
	]	
)
$, $
\lambda
=
\Eig[
	\frac{1}{
		\mathtt{w}^{-, 2, t}_{\Sf{x}}
	}
	\bs{M}_{\Sf{a} \Bs 1}^{+, 2, t}
	(\bs{M}_{\Sf{a} \Bs 1}^{+, 2, t})^{\Cts}
	+
	\frac{1}{
		\mathtt{w}^{-, 1, t}_{\Sf{x}}
	}
	\bs{M}_{\Sf{a}}^{+, 1, t}
	(\bs{M}_{\Sf{a}}^{+, 1, t})^{\Cts}
]
$ \\
$
\mathtt{c}_{\Sf{x}}^{+, 2, t + 1}
=
\Mean_{
	m_{\Sf{x}}^{+, 1, t},
	m_{\Sf{x}}^{+, 2, t}
}(
	\Mean_{\lambda}[
		\frac{1}{
			\lambda
			+
			\frac{1}{
				\mathtt{v}^{+, t + 1}_{\tilde{\Sf{x}}}
			}
		}
	]	
)
$, $
\lambda
=
\Eig[
	\frac{1}{
		\mathtt{w}^{-, 2, t}_{\Sf{x}}
	}
	\bs{M}_{\Sf{a}}^{+, 2, t}
	(\bs{M}_{\Sf{a}}^{+, 2, t})^{\Cts}
	+
	\frac{1}{
		\mathtt{w}^{-, 1, t}_{\Sf{x}}
	}
	\bs{M}_{\Sf{a} \Bs 1}^{+, 1, t}
	(\bs{M}_{\Sf{a} \Bs 1}^{+, 1, t})^{\Cts}
]
$ \\
$
\bs{m}_{\Sf{a}, k}^{+, 1, t + 1}
=
\bs{A}_{k}
\bs{m}_{\Sf{x}, k}^{+, 1, t + 1}
$ \\
$
\bs{m}_{\Sf{a}, k}^{+, 2, t + 1}
=
\bs{A}_{k}^{\Cts}
\bs{m}_{\Sf{x}, k}^{+, 2, t + 1}
$ \\
$
\mathtt{v}_{\Sf{z}}^{+, t + 1}
=
\frac{1}{M}
\mathtt{c}^{+, 1, t + 1}_{\Sf{x}}
\mathtt{c}^{+, 2, t + 1}_{\Sf{x}}
\Trace(
	\sum_{i = 1}^{M}{}
	\bs{A}_{i} \bs{A}_{i}^{\Cts}
)
+
\frac{1}{M}
\mathtt{c}^{+, 1, t + 1}_{\Sf{x}}
\Trace\{
	\Mean_{
		m_{\Sf{a}}^{+, 2, t + 1}
	}[
		\bs{M}_{\Sf{a}}^{+, 2, t + 1}
		(\bs{M}_{\Sf{a}}^{+, 2, t + 1})^{\Cts}
	]
\}
+
\linebreak
\frac{1}{M}
\mathtt{c}^{+, 2, t + 1}_{\Sf{x}}
\Trace\{
	\Mean_{
		m_{\Sf{a}}^{+, 1, t + 1}
	}[
		\bs{M}_{\Sf{a}}^{+, 1, t + 1}
		(\bs{M}_{\Sf{a}}^{+, 1, t + 1})^{\Cts}
	]
\}
$ \\
End \\
Output:
$\hat{\mathtt{v}}^{+, T + 1}_{\tilde{\Sf{x}}}$
\end{algorithmic}
\end{breakablealgorithm}

\section{Validation and Discussion}\label{Sec:Valid_Diss}
This section validates the QMP algorithm using two applications from engineering fields, i.e., compressed sensing for image processing and MIMO detection for wireless communications.
We consider the MMSE optimal estimator (\ref{Eq:MMSE_estimator}), i.e.,
$\beta \to 1$.
$\{\bs{A}_{i}\}_{i = 1, \cdots, M}$
is drawn from a Gaussian ensemble having i.i.d., zero-mean, and $\frac{1}{N}$-variance elements.

\subsection{Compressed Sensing}

In the application to compressed sensing, the generation process is specialized as
\begin{align*}
p(y | z)
= &
\NormalC[y | z, v_{\Sf{w}}]
.
\end{align*}
Throughout this part, we fix the parameters as $T = 30$, $N = 256$, and $M = 4 N$.

To allow denoisers have unknown explicit input-output relation, a QMP-based framework was proposed, and it is extended here to enhance QMP.
Following the convention there, this new algorithm is called D-QMP.
D-QMP is able to adopt denoisers that have no explicitly known input-output relation.
In the part, some prevailing denoisers of this type, called BM3D \cite{dabov2007image} and DnCNN \cite{zhang2017beyond}, are adopted, and the resulting new algorithms are termed BM3D-QMP and DnCNN-QMP, respectively.

\begin{figure}[!t]
\centering
\subfigure[Per-iteration MSEs at $v_{\Sf{w}} = \frac{1}{255}$]{
\label{Fig:0039216_i7_Iter}
\begin{minipage}[b]{0.4\linewidth}
\centering
\includegraphics[scale=0.3]{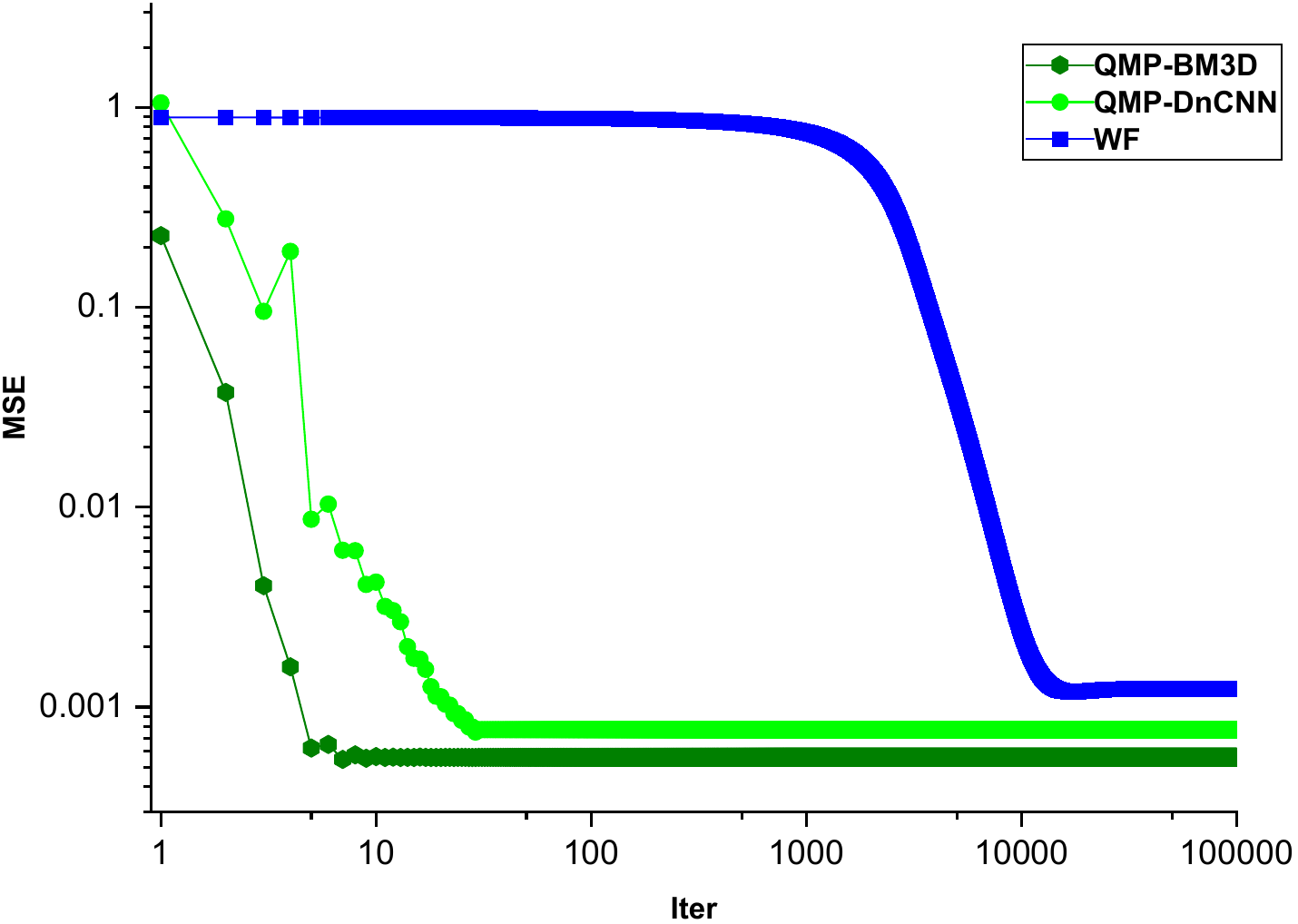}
\end{minipage}
}
\subfigure[Per-iteration MSEs at $v_{\Sf{w}} = \frac{10}{255}$]{
\label{Fig:039216_i7_Iter}
\begin{minipage}[b]{0.4\linewidth}
\centering
\includegraphics[scale=0.3]{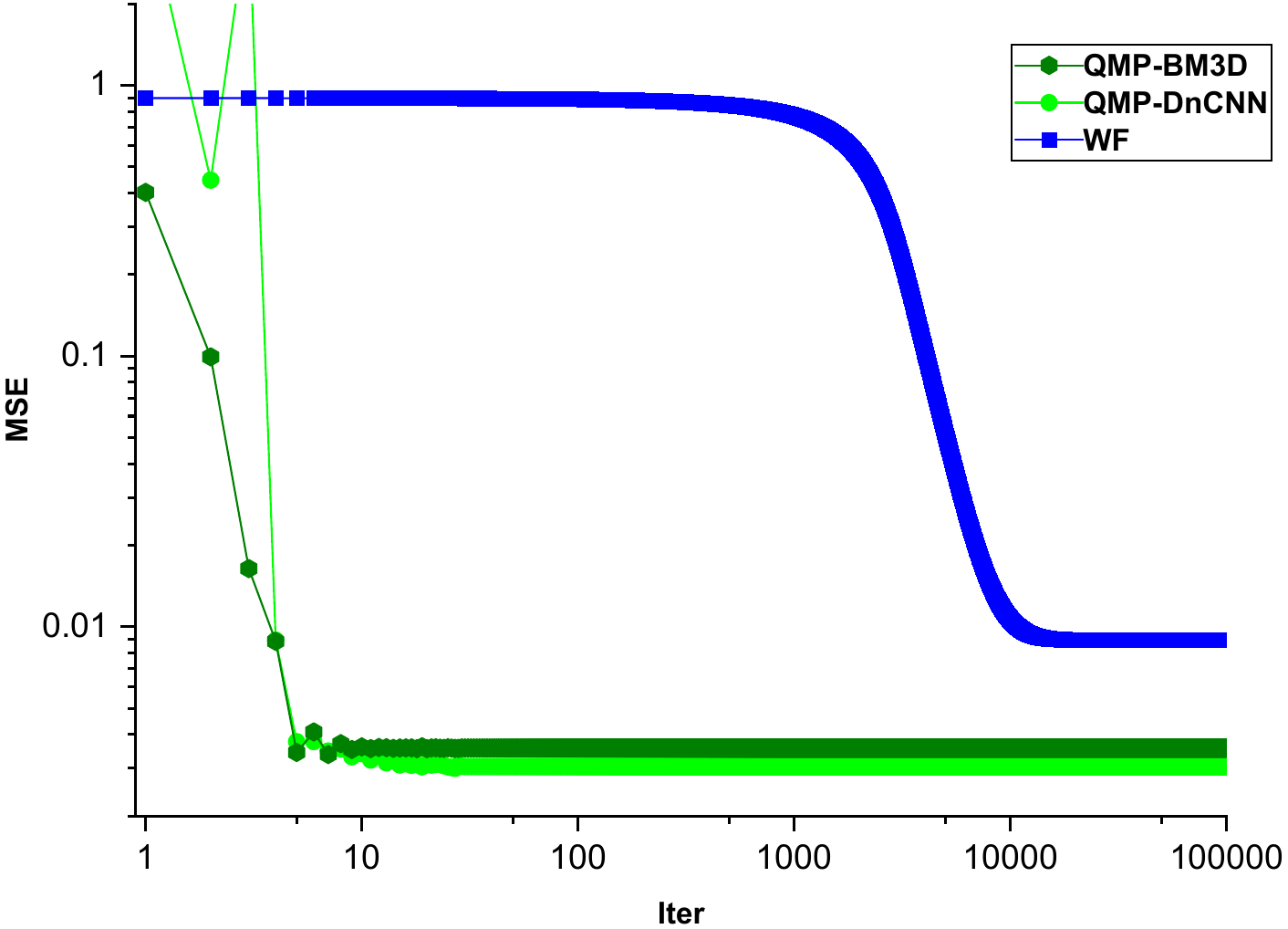}
\end{minipage}
}
\caption{
Per-iteration MSEs of WF and QMP.
}
\label{Fig:i7_Iter}
\end{figure}

\begin{figure}
\centering
\includegraphics[width=0.7\textwidth]{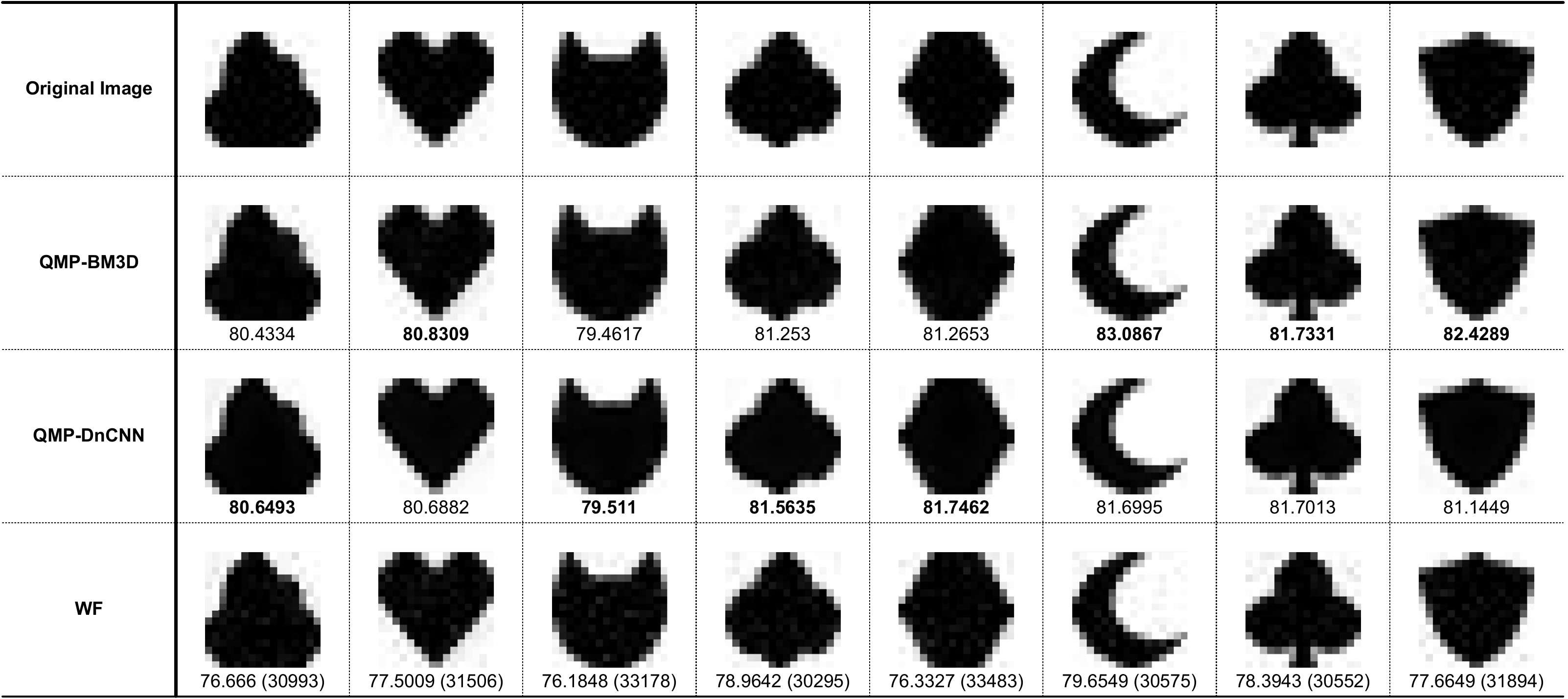}
\caption{
Denoising results for each image at
$v_{\Sf{w}} = \frac{1}{255}$.
}
\label{Fig:1_255_Index}
\end{figure}

\begin{figure}
\centering
\includegraphics[width=0.7\textwidth]{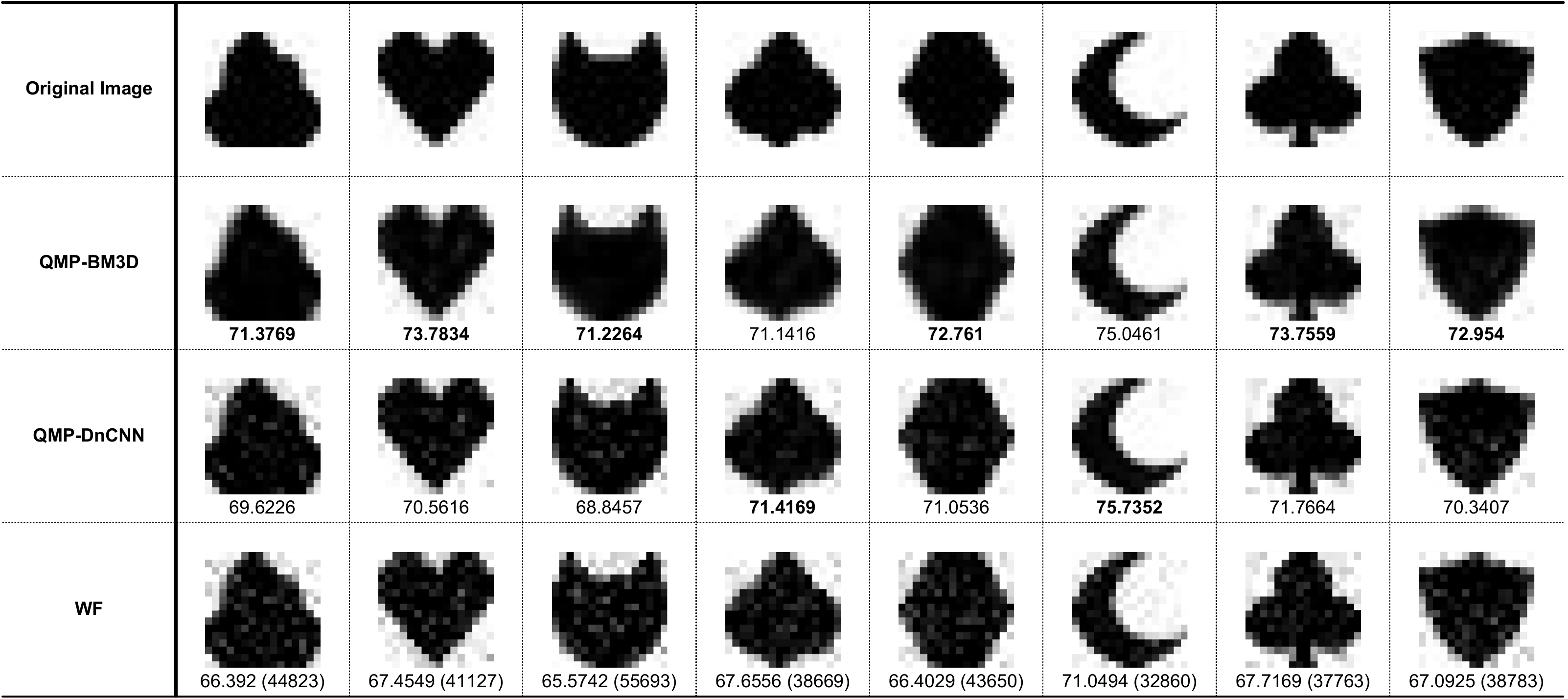}
\caption{
Denoising results for each image at
$v_{\Sf{w}} = \frac{10}{255}$.
}
\label{Fig:10_255_Index}
\end{figure}

Then, we have these remarks:

(i) Per-iteration behavior of QMP: 
We compare the MSE of QMP with Wirtinger flow (WF).
As shown in Fig. \ref{Fig:i7_Iter}, BM3D-QMP and DnCNN-QMP are highly effective:
the QMP-based algorithm converges in only a few iterations, which is much faster than WF, furthermore the fixed point of QMP is better.
For more evidences in the recovery of other images, Fig. \ref{Fig:1_255_Index} and \ref{Fig:10_255_Index} here provide an overall summary.

\begin{figure}
\centering
\includegraphics[width=0.7\textwidth]{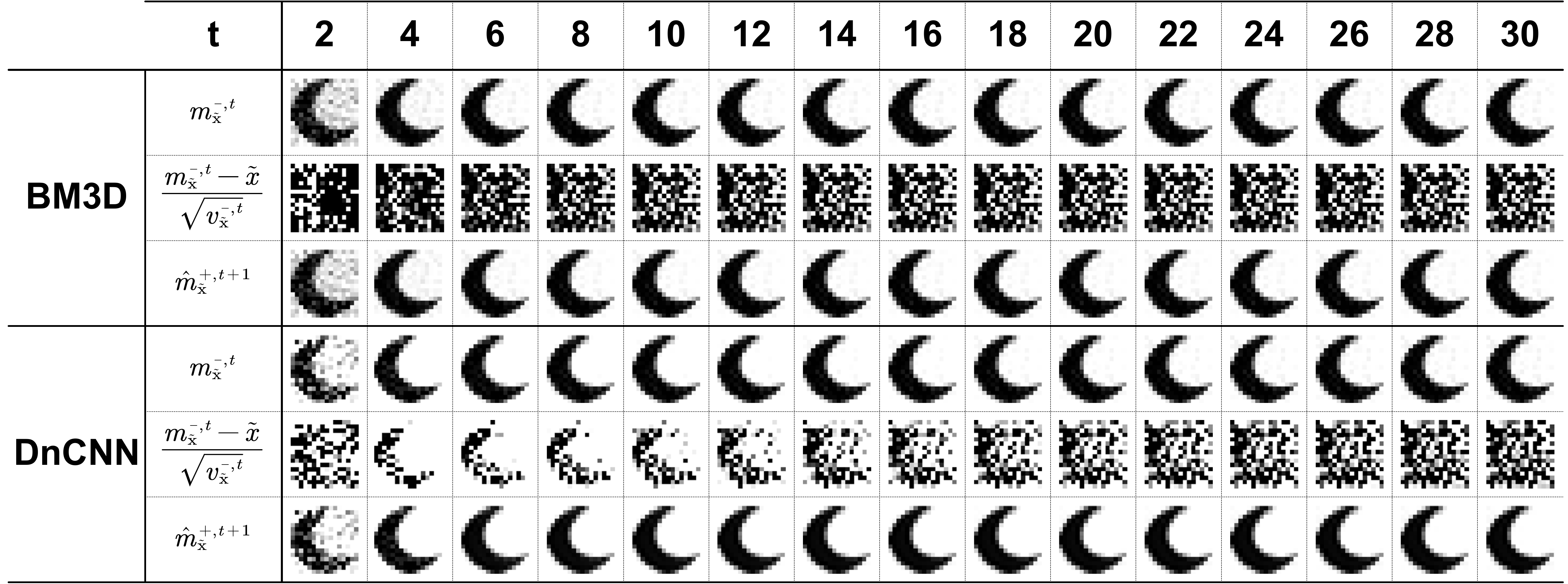}
\caption{
Per-iteration Microscopic variables at
$v_{\Sf{w}} = \frac{1}{255}$.
}
\label{Fig:1_255_Iteration}
\end{figure}

\begin{figure}
\centering
\includegraphics[width=0.7\textwidth]{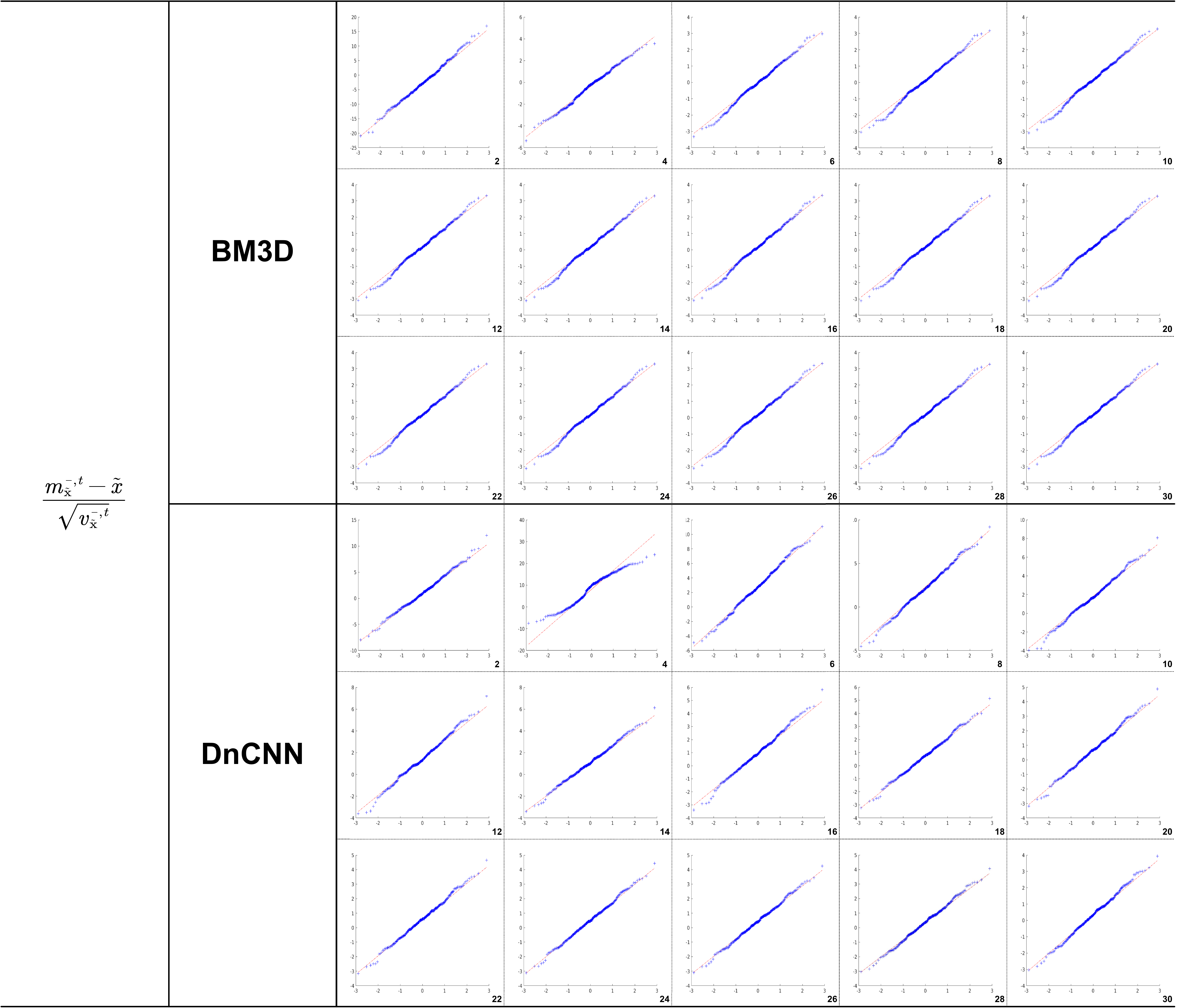}
\caption{
Microscopic analysis at
$v_{\Sf{w}} = \frac{1}{255}$
using QQ-plots (if two compared distributions are similar,
the points will approximately lie on the identity line)
}
\label{Fig:1_255_QQplot}
\end{figure}

\begin{figure}
\centering
\includegraphics[width=0.7\textwidth]{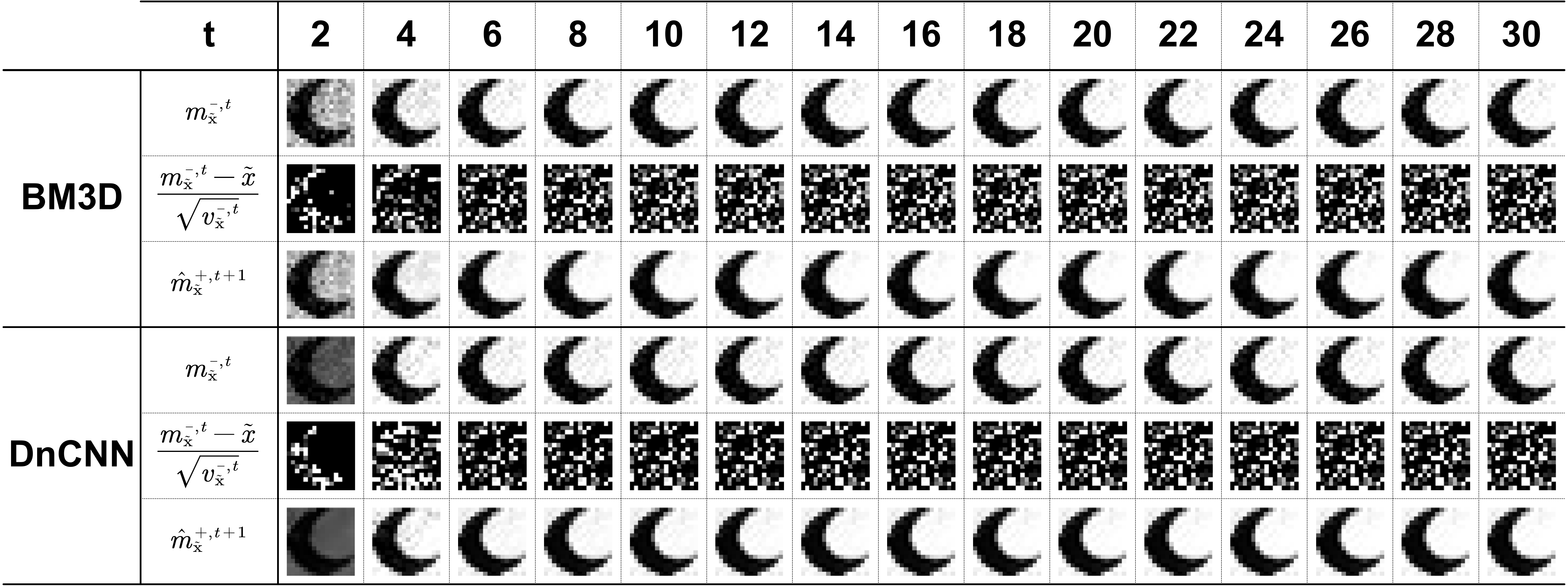}
\caption{
Per-iteration Microscopic variables at
$v_{\Sf{w}} = \frac{10}{255}$.
}
\label{Fig:10_255_Iteration}
\end{figure}

\begin{figure}
\centering
\includegraphics[width=0.7\textwidth]{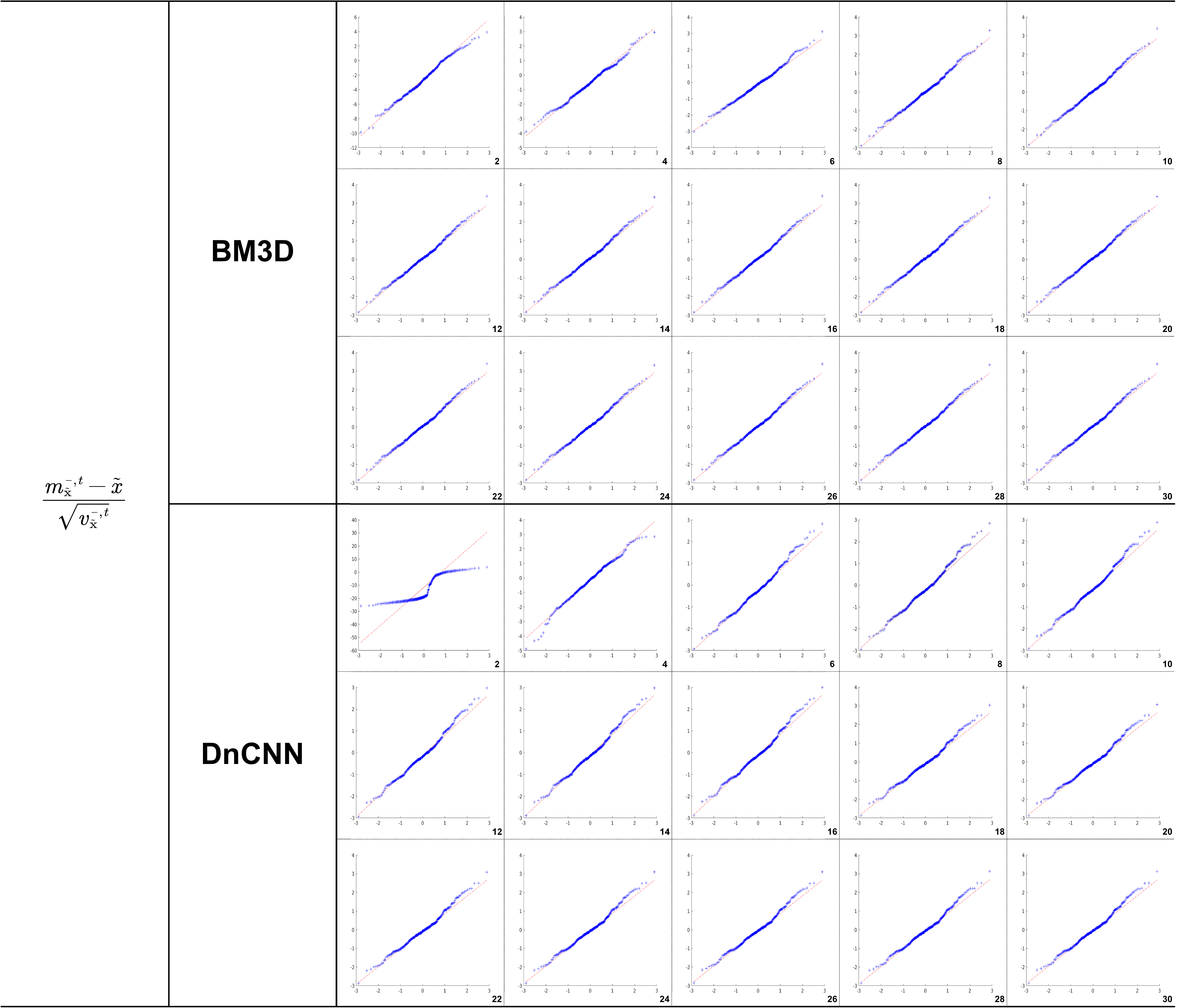}
\caption{
Microscopic analysis at
$v_{\Sf{w}} = \frac{10}{255}$
using QQ-plots (if two compared distributions are similar,
the points will approximately lie on the identity line)
}
\label{Fig:10_255_QQplot}
\end{figure}

(ii) Microscopic analysis:
As shown in Fig. \ref{Fig:1_255_Iteration}-\ref{Fig:1_255_QQplot} and \ref{Fig:10_255_Iteration}-\ref{Fig:10_255_QQplot},
a closer look into the Microscopic variables of the QMP-based algorithm, including
$\bs{m}_{\tilde{\Sf{x}}}^{-, t}$,
$
\frac{
	\bs{m}_{\tilde{\Sf{x}}}^{-, t} - \tilde{\bs{x}}
}{
	\sqrt{v_{\tilde{\Sf{x}}}^{-, t}}
}
$,
and
$\hat{\bs{m}}_{\tilde{\Sf{x}}}^{+, t + 1}$, presents the empirical convergence of QMP.

\subsection{MIMO Detection}

\begin{figure}[!t]
\centering
\subfigure[
0-1 case: Per-iteration MSEs at $v_{\Sf{w}} = 0.1, \rho = 0.55$
]{
\label{Fig:Exper_0_1_Iter}
\begin{minipage}[b]{0.4\linewidth}
\centering
\includegraphics[scale=0.3]{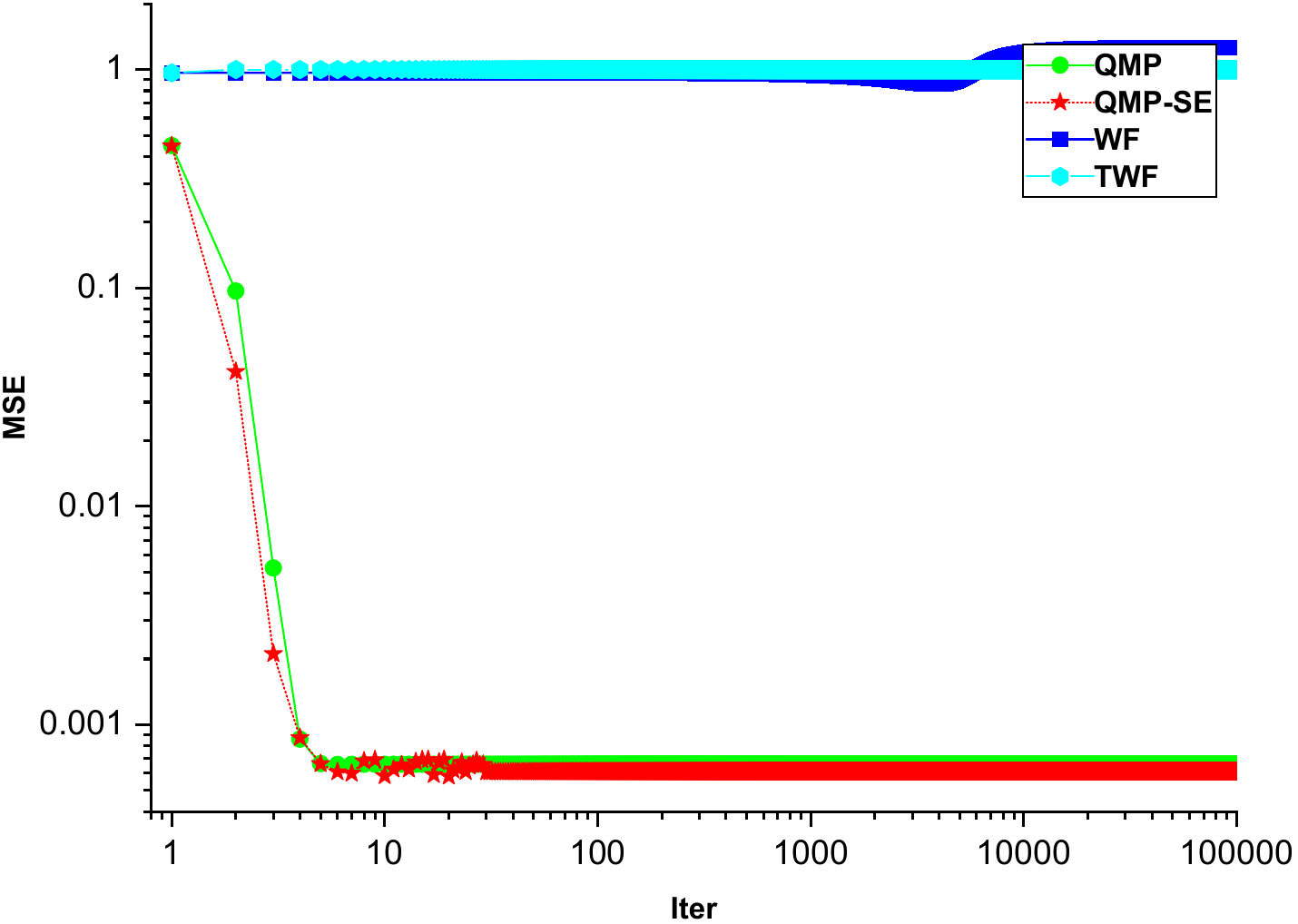}
\end{minipage}
}
\subfigure[
Uniform case: Per-iteration MSEs at $v_{\Sf{w}} = 0.05, a = 0.1, b = 2.1$
]{
\label{Fig:Exper_Uniform_Iter}
\begin{minipage}[b]{0.4\linewidth}
\centering
\includegraphics[scale=0.3]{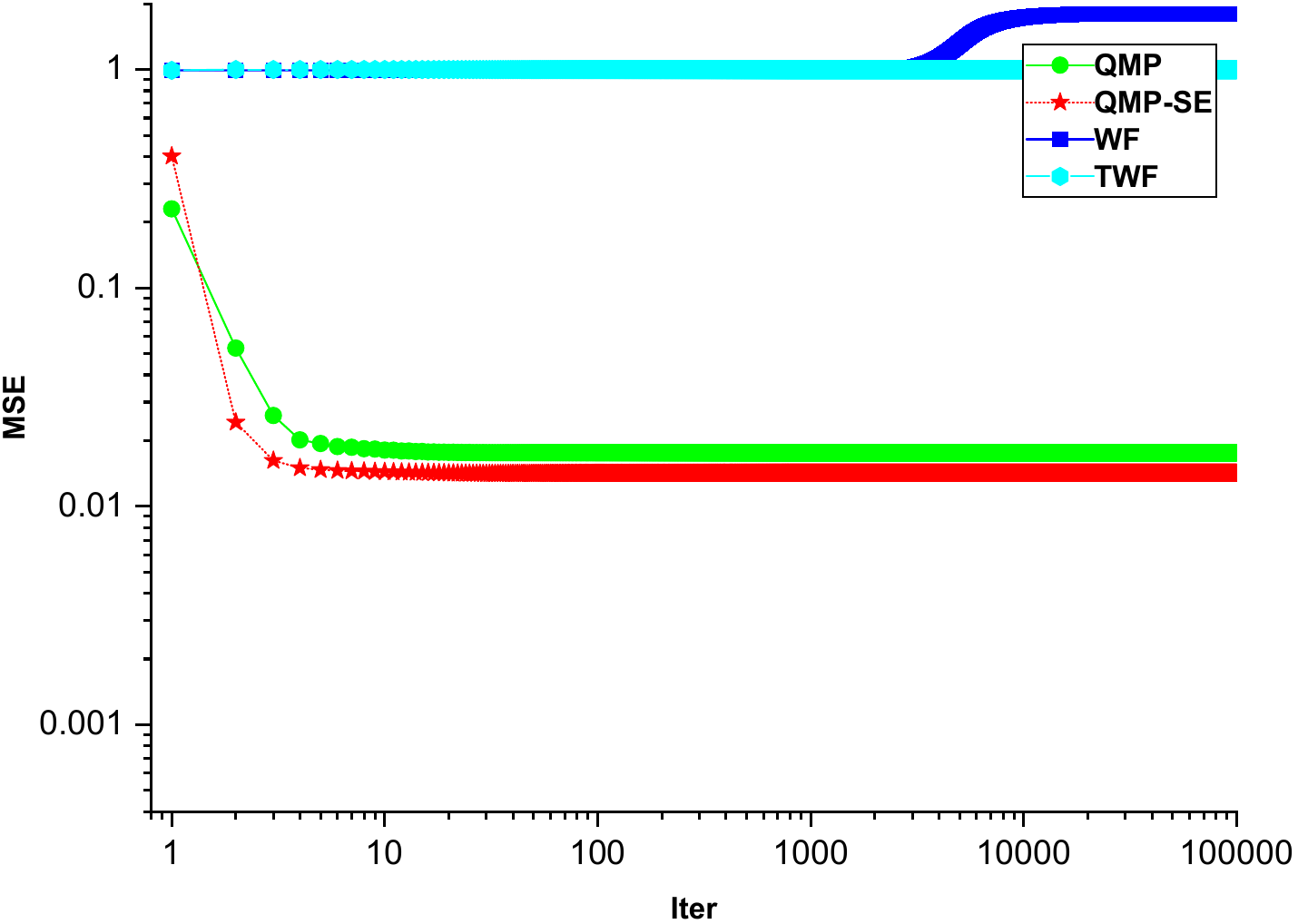}
\end{minipage}
}
\caption{
Per-iteration MSEs of WF, TWF, and QMP.
}
\label{Fig:Exper_Analytic_Prior}
\end{figure}

We apply the cases in the application to MIMO detection as follows:
\begin{itemize}
\item \TB{0-1 case}:
The generation process is specialized as
\begin{align*}
p(\tilde{x})
= &
(1 - \rho) \delta(\tilde{x})
+
\rho \delta(\tilde{x} - 1)
,\\
p(y | z)
= &
\Normal[y | z, v_{\Sf{w}}]
.
\end{align*}
Then the Line 13 of Algo. \ref{Algo:QMP} can be written as
\begin{align*}
\hat{m}_{\tilde{\Sf{x}}}^{+, t + 1}
= &
\frac{1}{C}
\int{\dd \tilde{x}}\,
\tilde{x}
p(\tilde{x})
\Normal[
	\tilde{x} |
	m_{\tilde{\Sf{x}}}^{-, t},
	v_{\tilde{\Sf{x}}}^{-, t}
]
=
\frac{1}{C}
\rho
\Normal[
	1 |
	m_{\tilde{\Sf{x}}}^{-, t},
	v_{\tilde{\Sf{x}}}^{-, t}
]
,\\
\hat{v}_{\tilde{\Sf{x}}}^{+, t + 1}
= &
\frac{1}{C}
\int{\dd \tilde{x}}\,
\tilde{x}^{2}
p(\tilde{x})
\Normal[
	\tilde{x} |
	m_{\tilde{\Sf{x}}}^{-, t},
	v_{\tilde{\Sf{x}}}^{-, t}
]
-
(\hat{m}_{\tilde{\Sf{x}}}^{+, t + 1})^{2}
=
\hat{m}_{\tilde{\Sf{x}}}^{+, t + 1}
-
(\hat{m}_{\tilde{\Sf{x}}}^{+, t + 1})^{2}
,
\end{align*}
with
\begin{align*}
C
\triangleq &
\int{\dd \tilde{x}}\,
p(\tilde{x})
\Normal[
	\tilde{x} |
	m_{\tilde{\Sf{x}}}^{-, t},
	v_{\tilde{\Sf{x}}}^{-, t}
]
=
(1 - \rho)
\Normal[
	0 |
	m_{\tilde{\Sf{x}}}^{-, t},
	v_{\tilde{\Sf{x}}}^{-, t}
]
+
\rho
\Normal[
	1 |
	m_{\tilde{\Sf{x}}}^{-, t},
	v_{\tilde{\Sf{x}}}^{-, t}
]
.
\end{align*}
And the Line 16 of Algo. \ref{Algo:SE:QS_MP} can be denoted as
\begin{align*}
q^{+, t + 1}_{\tilde{\Sf{x}}}
= &
\int{\dd \zeta}\,
\frac{
	(
		\int{\dd \tilde{x}}\,
		\tilde{x} p(\tilde{x})
		\Normal[
			\tilde{x} | \zeta,
			\mathtt{v}^{-, t}_{\tilde{\Sf{x}}}
		]
	)^{2}
}{
	\int{\dd \tilde{x}}\,
	p(\tilde{x})
	\Normal[
		\tilde{x} | \zeta,
		\mathtt{v}^{-, t}_{\tilde{\Sf{x}}}
	]
}
=
\int{\dd u}\,
\frac{
	q^{2}
	\EXP[
		\frac{u}{
			\sqrt{v_{\tilde{\Sf{x}}}^{-, t}}
		}
		-
		\frac{1}{
			2 v_{\tilde{\Sf{x}}}^{-, t}
		}
	]
}{
	(1 - q)
	\EXP[
		\frac{1}{
			2 v_{\tilde{\Sf{x}}}^{-, t}
		}
		-
		\frac{u}{
			\sqrt{v_{\tilde{\Sf{x}}}^{-, t}}
		}
	]
	+
	q
}
\Normal[u | 0, 1]
;
\end{align*}

\item \TB{uniform case}:
The generation process is specialized as
\begin{align*}
p(\tilde{x})
= &
\begin{cases}
\frac{1}{b - a} & \text{for }a \leq x \leq b \\
0 & \text{for } x < a \text{ or } x > b 
\end{cases}
,\\
p(y | z)
= &
\Normal[y | z, v_{\Sf{w}}]
.
\end{align*}
Then the Line 13 of Algo. \ref{Algo:QMP} can be written as
\begin{align*}
\hat{m}_{\tilde{\Sf{x}}}^{+, t + 1}
\triangleq &
\frac{1}{C}
\int_{a}^{b}{\dd \tilde{x}}\,
\tilde{x}
p(\tilde{x})
\Normal[
	\tilde{x} |
	m_{\tilde{\Sf{x}}}^{-, t},
	v_{\tilde{\Sf{x}}}^{-, t}
]
=
m_{\tilde{\Sf{x}}}^{-, t}
+
\frac{
	\phi(\alpha) - \phi(\beta)
}{C}
\sqrt{
	v_{\tilde{\Sf{x}}}^{-, t}
}
,\\
\hat{v}_{\tilde{\Sf{x}}}^{+, t + 1}
\triangleq &
\frac{1}{C}
\int_{a}^{b}{\dd \tilde{x}}\,
(
	\tilde{x}
	-
	\hat{m}_{\tilde{\Sf{x}}}^{+, t + 1}
)^{2}
p(\tilde{x})
\Normal[
	\tilde{x} |
	m_{\tilde{\Sf{x}}}^{-, t},
	v_{\tilde{\Sf{x}}}^{-, t}
]
=
v_{\tilde{\Sf{x}}}^{-, t}
[
	1
	+
	\frac{
		\alpha \phi(\alpha)
		-
		\beta \phi(\beta)
	}{C}
	-
	(
		\frac{
			\phi(\alpha) - \phi(\beta)
		}{C}
	)^{2}
]
,
\end{align*}
with
\begin{align*}
\alpha
\triangleq &
\frac{
	a - m_{\tilde{\Sf{x}}}^{-, t}
}{
	\sqrt{
		v_{\tilde{\Sf{x}}}^{-, t}
	}
}
,\\
\beta
\triangleq &
\frac{
	b - m_{\tilde{\Sf{x}}}^{-, t}
}{
	\sqrt{
		v_{\tilde{\Sf{x}}}^{-, t}
	}
}
,\\
C
\triangleq &
\int_{a}^{b}{\dd \tilde{x}}\,
p(\tilde{x})
\Normal[
	\tilde{x} |
	m_{\tilde{\Sf{x}}}^{-, t},
	v_{\tilde{\Sf{x}}}^{-, t}
]
=
\Phi(\beta) - \Phi(\alpha)
,
\end{align*}
with $
\phi(\cdot)
$ and $
\Phi(\cdot)
$ being the standard normal distribution and its cumulative distribution function, respectively.
And the Line 17 of Algo. \ref{Algo:SE:QS_MP} can be denoted as
\begin{align*}
\hat{\mathtt{v}}^{+, t + 1}_{\tilde{\Sf{x}}}
= &
\int{\dd m_{\tilde{\Sf{x}}}^{-, t}}
\int_{a}^{b}{\dd \tilde{x}}\,
p(m_{\tilde{\Sf{x}}}^{-, t}, \tilde{x})
\hat{v}_{\tilde{\Sf{x}}}^{+, t + 1}(m_{\tilde{\Sf{x}}}^{-, t})
\\
= &
\int{\dd m_{\tilde{\Sf{x}}}^{-, t}}
\int_{a}^{b}{\dd \tilde{x}}\,
p(m_{\tilde{\Sf{x}}}^{-, t} | \tilde{x}) p(\tilde{x})
\hat{v}_{\tilde{\Sf{x}}}^{+, t + 1}(m_{\tilde{\Sf{x}}}^{-, t})
\\
= &
\frac{1}{b - a}
\int{\dd m_{\tilde{\Sf{x}}}^{-, t}}\,
(
	\int_{a}^{b}{\dd \tilde{x}}\,
	\Normal[
		\tilde{x} |
		m_{\tilde{\Sf{x}}}^{-, t},
		v_{\tilde{\Sf{x}}}^{-, t}
	]
)
\hat{v}_{\tilde{\Sf{x}}}^{+, t + 1}(m_{\tilde{\Sf{x}}}^{-, t})
\\
= &
\frac{1}{b - a}
\int{\dd m_{\tilde{\Sf{x}}}^{-, t}}\,
C(m_{\tilde{\Sf{x}}}^{-, t})
\hat{v}_{\tilde{\Sf{x}}}^{+, t + 1}(m_{\tilde{\Sf{x}}}^{-, t})
.
\end{align*}

\end{itemize}

Throughout this subsection, we fix the parameters as $T = 30$, $N = 256$, and $M = 4 N$.
We compare the MSE of QMP with Wirtinger flow (WF) andthe thresholded Wirtinger flow (TWF) algorithm.
As shown in Fig. \ref{Fig:Exper_0_1_Iter} and \ref{Fig:Exper_Uniform_Iter}, QMP is highly effective:
QMP outperforms WF and TWF significantly.
QMP converges in only a few iterations.
And the SE of QMP is observed to capture the dynamical behavior of QMP perfectly.

\section{Conclusion}

For approximate inference in the generalized quadratic equations model, many state-of-the-art algorithms lack any prior knowledge of the target signal structure, exhibits slow convergence, and can not handle any analytic prior knowledge of the target signal structure.
So, we introduced the QMP algorithm in this paper.
The QMP obtained is as computationally efficient.
We also derived the SE for QMP, which can capture the per-iteration behavior of the simulated algorithm precisely.
To see more evidence on our theoretical findings, we carried out extensive simulations and confirmed that QMP generally outperforms other state-of-the-art inference methods.

%
%
%
%

\ifCLASSOPTIONcaptionsoff
  \newpage
\fi



\begin{thebibliography}{10}
\providecommand{\url}[1]{#1}
\csname url@samestyle\endcsname
\providecommand{\newblock}{\relax}
\providecommand{\bibinfo}[2]{#2}
\providecommand{\BIBentrySTDinterwordspacing}{\spaceskip=0pt\relax}
\providecommand{\BIBentryALTinterwordstretchfactor}{4}
\providecommand{\BIBentryALTinterwordspacing}{\spaceskip=\fontdimen2\font plus
\BIBentryALTinterwordstretchfactor\fontdimen3\font minus
  \fontdimen4\font\relax}
\providecommand{\BIBforeignlanguage}[2]{{%
\expandafter\ifx\csname l@#1\endcsname\relax
\typeout{** WARNING: IEEEtran.bst: No hyphenation pattern has been}%
\typeout{** loaded for the language `#1'. Using the pattern for}%
\typeout{** the default language instead.}%
\else
\language=\csname l@#1\endcsname
\fi
#2}}
\providecommand{\BIBdecl}{\relax}
\BIBdecl

\bibitem{shechtman2011sparsity}
Y.~Shechtman, Y.~C. Eldar, A.~Szameit, and M.~Segev, ``Sparsity based
  sub-wavelength imaging with partially incoherent light via quadratic
  compressed sensing,'' \emph{Optics express}, vol.~19, no.~16, pp.
  14\,807--14\,822, 2011.

\bibitem{helmberg1998solving}
C.~Helmberg and F.~Rendl, ``Solving quadratic (0, 1)-problems by semidefinite
  programs and cutting planes,'' \emph{Mathematical programming}, vol.~82, pp.
  291--315, 1998.

\bibitem{gerchberg1994practical}
R.~Gerchberg and W.~Saxton, ``A practical algorithm for the determination of
  phase from image and diffraction plane pictures,'' \emph{SPIE milestone
  series MS}, vol.~93, pp. 306--306, 1994.

\bibitem{fienup1982phase}
J.~R. Fienup, ``Phase retrieval algorithms: a comparison,'' \emph{Applied
  optics}, vol.~21, no.~15, pp. 2758--2769, 1982.

\bibitem{candes2013phaselift}
E.~J. Candes, T.~Strohmer, and V.~Voroninski, ``Phaselift: Exact and stable
  signal recovery from magnitude measurements via convex programming,''
  \emph{Communications on Pure and Applied Mathematics}, vol.~66, no.~8, pp.
  1241--1274, 2013.

\bibitem{candes2014solving}
E.~J. Cand{\`e}s and X.~Li, ``Solving quadratic equations via phaselift when
  there are about as many equations as unknowns,'' \emph{Foundations of
  Computational Mathematics}, vol.~14, pp. 1017--1026, 2014.

\bibitem{bishop2006pattern}
C.~M. Bishop and N.~M. Nasrabadi, \emph{Pattern recognition and machine
  learning}.\hskip 1em plus 0.5em minus 0.4em\relax Springer, 2006, vol.~4,
  no.~4.

\bibitem{mezard2009information}
M.~Mezard and A.~Montanari, \emph{Information, physics, and computation}.\hskip
  1em plus 0.5em minus 0.4em\relax Oxford University Press, 2009.

\bibitem{rasmussen2003gaussian}
C.~E. Rasmussen, ``Gaussian processes in machine learning,'' in \emph{Summer
  school on machine learning}.\hskip 1em plus 0.5em minus 0.4em\relax Springer,
  2003, pp. 63--71.

\bibitem{rubinstein2016simulation}
R.~Y. Rubinstein and D.~P. Kroese, \emph{Simulation and the {Monte} {Carlo}
  method}.\hskip 1em plus 0.5em minus 0.4em\relax John Wiley \& Sons, 2016.

\bibitem{metropolis1949monte}
N.~Metropolis and S.~Ulam, ``The monte carlo method,'' \emph{Journal of the
  American statistical association}, vol.~44, no. 247, pp. 335--341, 1949.

\bibitem{dabov2007image}
K.~Dabov, A.~Foi, V.~Katkovnik, and K.~Egiazarian, ``Image denoising by sparse
  {3-D} transform-domain collaborative filtering,'' \emph{{IEEE} Trans. Image
  Process.}, vol.~16, no.~8, pp. 2080--2095, 2007.

\bibitem{zhang2017beyond}
K.~Zhang, W.~Zuo, Y.~Chen, D.~Meng, and L.~Zhang, ``Beyond a gaussian denoiser:
  {Residual} learning of deep cnn for image denoising,'' \emph{{IEEE} Trans.
  Image Process.}, vol.~26, no.~7, pp. 3142--3155, 2017.

\end{thebibliography}
\end{document}